\documentclass[12pt]{article}
\usepackage{newinutile}
\usepackage{amsmath,amssymb}
\usepackage{epsfig,graphics,color,calc,graphicx}
\usepackage{mathrsfs}

\newcommand{\rr}[1]{{\normalfont\textrm{#1}}}

\begin{document}
\title{Anticipation decides  on lane formation in pedestrian
counterflow -- a simulation study}

\author{Emilio N.M.\ Cirillo}
\email{emilio.cirillo@uniroma1.it}
\affiliation{Dipartimento di Scienze di Base e Applicate per 
             l'Ingegneria, Sapienza Universit\`a di Roma, 
             via A.\ Scarpa 16, I--00161, Roma, Italy.}

\author{Adrian Muntean}
\email{adrian.muntean@kau.se}
\affiliation{Department of Mathematics and Computer Science, 
Karlstad University, Sweden.}


\begin{abstract}
Human crowds base most of their behavioral decisions upon anticipated states of their walking environment. We explore a minimal version of a 
lattice model to study  lanes formation in pedestrian counterflow. 
Using the concept of horizon depth, our simulation results suggest that the anticipation effect together with the presence of a small background noise play an important role in promoting collective behaviors in a counterflow setup. These ingredients facilitate the formation of seemingly stable lanes and ensure the ergodicity of the system.
\end{abstract}



\preprint{Appunti: \today}


\maketitle


\section{Introduction}
\label{s:lane}
\subsection{Lanes in counterflow}
Very much like colloids or bacteria colonies, human crowds can be thought of as  many--particle interacting 
systems. From this perspective, the non--equilibrium statistical mechanics becomes the right language to study large crowds coming into play, where highly complex situations giving rise to interesting collective 
phenomena, such as free flow, lanes\footnote{One of the most efficient transport mechanisms in crowded situations -- like those where active colloidal particles  are supposed to cross soft matter solutions -- or when pedestrian counterflow in Tokyo's  Shibuya and Shinjuku railways stations  thrives for fluidization -- are lanes.  An average of 3.5 million people per day use the Shinjuku station, making it the busiest station in the world in terms of passenger numbers. Shibuya station is similarly busy.},
and gridlock or jamming (cf. \cite{Bellomo,Doi13,Sum06,ST19}, e.g.), may arise. 

This paper focuses on \emph{lane formation} in pedestrian 
counterflow, i.e.,
a bidirectional pedestrian movement. The novelty we bring into this context is linked to the ambition to reproduce the rational behaviour of individuals forming lanes through the means of anticipation, an old idea that can be traced back at least from Oresme's time; see \cite{Oreste}. From this perspective, we are in line with some of the statements in \cite{Bailo} and complement existing research on the lane formation topic in pedestrian counterflows.  

\subsection{A brief literature review on lane formation}

The mechanisms yielding lane formation are still obscure and object of current research from both theoretical and experimental viewpoints. From the crowd management perspectives, one strongly believes that controlling in real time the building up and the dissolution of lanes would be an efficient tool  both for organization matters as well as for what  concerns the activity of responsible law enforcement agencies.  Regarding lane formation, we refer  the reader to \cite{NS} for a review from the point of view of self--organization, to \cite{HKHE12} for a perspective 
from the transportation engineering side, as well as to own previous research \cite{Joep} where we attempted to investigate pedestrian counterflows through heterogeneous domains. 
Lane formation has been observed in many empirical studies (see 
the discussion in \cite[Section~4.1]{HFMV}) and it has also 
been noted that a certain degree of noise favors lane 
formation, whereas a too large noise amplitude lead to a 
``freezing by heating'' effect \cite{HP,HV}. It is worth also looking into the recent study \cite{MFN19}, where lanes are perceived as 
super--diffusive L\'evy walks.  The formation of lanes has been quite well described in the framework of the 
\emph{social force model} \cite{H1991,HFMV}, whereas it has been considered 
a quite hard phenomenon to be described in the framework of more 
elementary \emph{cellular automata} models
\cite{CM12,CM13}.  Related ideas are reported, for instance,  also in \cite{Degond0, Degond1}. The main 
drawback of the social force model is that it involves a large number of parameters. An important breakthrough in this 
direction can be considered the paper \cite{BKSZ} in which 
the idea of the \emph{floor field} cellular automaton 
has been firstly introduced. 
In this model the floor field is constructed dynamically during 
the evolution of the system and allows the coupling between 
the motion of the particles and a sort of \emph{trace} left
by particles which moved before \cite{NS,SYN}. 
The floor field is traditionally made of a \emph{static} and a
\emph{dynamic} component \cite{BKSZ}. More recently 
a so called \emph{anticipation} component has also 
been taken into account \cite{NS,SYN}. We refer the reader to \cite{GTLW17} for an account of anticipation effects in the context of  deterministic dynamical system modeling pedestrian motion. 
The static floor field is constant in time and not influenced 
by the presence of other particles, it simply codes the 
preferential direction of motion of each particle. 
The dynamic floor field, inspired by the motion of ants who
leave pheromone traces which can be smelled by other ants, 
evolve with time and codes the trace left by moving pedestrians.
The anticipation floor field allows pedestrian to estimate 
the route of pedestrians moving in the opposite direction and try to 
avoid collisions. 

\subsection{Aim of this research}

In this framework, we propose an elementary model as well as 
a different mechanism for lane 
formation. The main idea behind this mechanism is mildly related to the 
anticipation floor field  just discussed previously.
The aim of our study is to bring evidence that the elementary mechanism 
yielding lane formation is the pedestrian's attitude to 
avoid collisions with pedestrians moving in the opposite direction, i.e. the {\em anticipation}. 

To keep as simple as possible the modeling level, we use a lattice model
approach. We define a discrete time dynamics
on a lattice with an exclusion rule, namely, each site can be 
occupied by a single particle at time. 
The formation of lanes at stationarity is studied 
by means of the order parameter proposed in \cite{NS}.  We demonstrate that,
provided the attitude to avoid collisions is relevant enough, 
lanes naturally appear in the system.  This is in our view the main mechanism leading to the formation of lanes.

The rest of the paper is organized as follows. 
In Section~\ref{s:aim} we describe in more detail the crowd dynamics scenarion we have in mind. The model is presented in Section~\ref{s:lane-mod}.
The results of our simulations are discussed in 
Section~\ref{s:lane-sim}, whereas our conclusions are finally summarized in 
Section~\ref{s:dis}.

\section{The model}
\label{s:mod-intro}

In this section,  we present the crowd dynamics scenario we have in mind. Here we define the chosen modeling strategy and briefly explain the main observables that will  be closely followed in our simulations. These observables  are our main tools to explore the internal coherent crowd structures which are expected to form  in pedestrian counterflows.

\subsection{Pedestrian counterflow as a Gedankenexperiment}
\label{s:aim}
Our crowd dynamics setup is as follows: Two different types of pedestrians enter a vertical 
strip: those moving upward (``red particles") and 
those moving downward (``blue particles").
At each time the pedestrians move mostly forward with respect to 
their preferential direction, but they will have a small 
probability $r$, called the \emph{background noise}, 
to do something different, namely, stepping laterally or 
even moving backward. 
This noise mimics the presence of irregularities (small obstacles) 
in the strip, or simply, 
the pedestrian's loss of visual focus due to 
interactions with the surrounding ambient as it is often promoted in environmental psychology reports. The reason for such a background noise is also technical. Indeed, 
the mathematical model turns to be a discrete time Markov Chain. 
In the case $r=0$, the model would exhibit many absorbing varieties made of those 
configurations in which particles are perfectly in--lane. In other words, 
set of configurations in which columns are occupied either by red or 
blue particles would be absorbing varieties of the state space. 
Considering $r>0$ simply ensures the  ergodicity\footnote{Ergodicity is lost for instance when an horizontal line with blue particles opposing red particles is formed.} of the model. 

Pedestrians do not move simultaneously, but sequentially;  namely, 
the new position reached by a particular pedestrian has to be 
taken into account when moving the following one. For this reason, we refer to
our model  as  \emph{lattice model} rather than cellular automaton.

The distinguishing feature we introduce in this context is the idea of 
\emph{horizon}: if one pedestrian spots another one moving 
in the opposite direction in front of him within an {\em a priori}  fixed 
distance (i.e., the horizon depth), 
then she/he will try to step laterally with 
the probability $h$.
The correlation between the lateral 
motion of a particle and the approaching of an opposing 
one will yield lane formation as we see in Section \ref{s:lane-sim}. 
The model will be explored by numerical simulations for different values of the 
parameters. Having in view the application to pedestrian motions,  
the relevant parameter regime is $r\ll 1$ (low noise background) and 
$h\gg0$ (large avoiding tendency). 

\subsection{Definition of the  model}
\label{s:lane-mod}
The model is very much inspired by the one proposed by the authors in 
\cite{CKMSS16,CKMS16}.
The walking space is chosen to be  the \emph{strip} 
$\Lambda=\{1,\dots,L_1\}\times\{1,\dots,L_2\}\subset\mathbb{Z}^2$.
Each \emph{site} or \emph{cell}
in $\Lambda$ can be either empty or occupied by a 
single particle (hard core repulsion). 

\setlength{\unitlength}{0.8pt}
\begin{figure}[h]
\begin{picture}(400,330)(-160,20)
\thinlines
\multiput(40,0)(20,0){8}{\line(0,1){300}}
\multiput(40,0)(0,20){16}{\line(1,0){140}}
\put(100,-45){\vector(-1,0){60}}
\put(120,-45){\vector(1,0){60}}
\put(104,-49){${L_1}$}
\put(200,140){\vector(0,-1){140}}
\put(200,160){\vector(0,1){140}}
\put(194,145){${L_2}$}
\linethickness{0.5mm}
\put(60,240){\begin{color}{red}\line(1,0){20}\end{color}}
\put(60,240){\begin{color}{red}\line(0,-1){60}\end{color}}
\put(80,240){\begin{color}{red}\line(0,-1){60}\end{color}}
\put(60,180){\begin{color}{red}\line(1,0){20}\end{color}}
\thinlines
\put(70,250){\vector(0,1){13}}
\put(70,250){\vector(0,-1){13}}
\put(70,250){\vector(1,0){13}}
\put(70,250){\vector(-1,0){13}}
\put(70,250){\begin{color}{red}\circle*{5}\end{color}}
\put(70,190){\begin{color}{red}\circle*{5}\end{color}}
\put(61,227){${\scriptstyle 1-3r/4}$}
\put(63,267){${\scriptstyle r/4}$}
\put(41,247){${\scriptstyle r/4}$}
\put(85,247){${\scriptstyle r/4}$}

\linethickness{0.5mm}
\put(140,260){\begin{color}{red}\line(1,0){20}\end{color}}
\put(140,260){\begin{color}{red}\line(0,-1){60}\end{color}}
\put(160,260){\begin{color}{red}\line(0,-1){60}\end{color}}
\put(140,200){\begin{color}{red}\line(1,0){20}\end{color}}
\thinlines
\put(150,270){\vector(0,1){13}}
\put(150,270){\vector(0,-1){13}}
\put(150,270){\vector(1,0){13}}
\put(150,270){\vector(-1,0){13}}
\put(150,270){\begin{color}{red}\circle*{5}\end{color}}
\put(150,210){\begin{color}{blue}\circle*{5}\end{color}}
\put(141,247){${\scriptstyle 1-h}$}
\put(148,287){${\scriptstyle 0}$}
\put(121,267){${\scriptstyle h/2}$}
\put(165,267){${\scriptstyle h/2}$}

\linethickness{0.5mm}
\put(80,40){\begin{color}{blue}\line(1,0){20}\end{color}}
\put(80,40){\begin{color}{blue}\line(0,1){60}\end{color}}
\put(100,40){\begin{color}{blue}\line(0,1){60}\end{color}}
\put(80,100){\begin{color}{blue}\line(1,0){20}\end{color}}
\thinlines
\put(90,30){\vector(0,1){13}}
\put(90,30){\vector(0,-1){13}}
\put(90,30){\vector(1,0){13}}
\put(90,30){\vector(-1,0){13}}
\put(90,30){\begin{color}{blue}\circle*{5}\end{color}}
\put(90,70){\begin{color}{blue}\circle*{5}\end{color}}
\put(81,47){${\scriptstyle 1-3r/4}$}
\put(83,7){${\scriptstyle r/4}$}
\put(61,27){${\scriptstyle r/4}$}
\put(105,27){${\scriptstyle r/4}$}

\linethickness{0.5mm}
\put(120,120){\begin{color}{blue}\line(1,0){20}\end{color}}
\put(120,120){\begin{color}{blue}\line(0,1){60}\end{color}}
\put(140,120){\begin{color}{blue}\line(0,1){60}\end{color}}
\put(120,180){\begin{color}{blue}\line(1,0){20}\end{color}}
\thinlines
\put(130,110){\vector(0,1){13}}
\put(130,110){\vector(0,-1){13}}
\put(130,110){\vector(1,0){13}}
\put(130,110){\vector(-1,0){13}}
\put(130,110){\begin{color}{blue}\circle*{5}\end{color}}
\put(130,150){\begin{color}{red}\circle*{5}\end{color}}
\put(121,127){${\scriptstyle 1-h}$}
\put(128,87){${\scriptstyle 0}$}
\put(101,107){${\scriptstyle h/2}$}
\put(145,107){${\scriptstyle h/2}$}
\end{picture}
\vskip 2. cm
\caption{Schematic representation of the model for 
horizon $H=3$. Arrows denote possible moves and the 
related probabilities are reported in the cell.
}
\label{f:reticolo}
\end{figure}
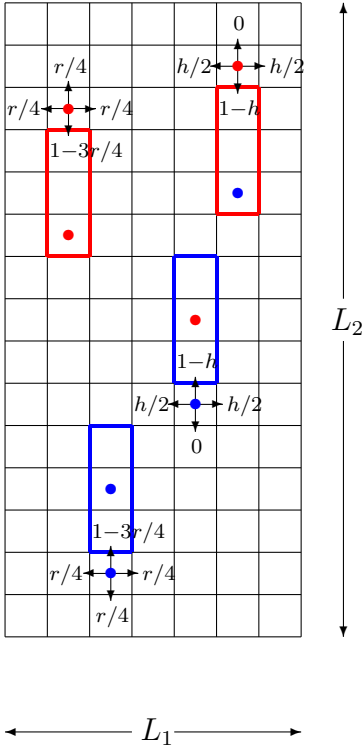
\setlength{\unitlength}{1.pt}

Each particle is either \emph{red} or \emph{blue}. 
For red particles, the \emph{forward} direction is downward, 
whereas for blue particles the forward direction is upward. 
We let $N_\rr{r}$ and $N_\rr{b}$ be the number of 
red and blue particles at the initial time $t=0$,  respectively.  
We let $N:=N_\rr{r}+N_\rr{b}$ be total number of particles at the initial 
time and $n(t)$ be the total number of particles at time $t$. 

Particles are labelled.
 At each step of the dynamics, we choose sequentially at random with uniform 
probability one of the $N$ particles. If the particle lies in the lattice, then 
we  displace it
with the probabilities specified below: if the cell 
where the particle should be moved to is occupied, then 
the particle is not moved (hard core repulsion acts accordingly to the simple exclusion rule). 
Time is increased by one after $N$ particles have been selected and 
possibly  moved. 

We let $r\in[0,1]$ be the \emph{background noise}
and
$h\in[0,1]$ be the \emph{lateral move probability}. 
Moreover,
given a particle, its \emph{horizon} is the vertical slab 
made of 
the first $H\ge0$ cells 
(up or down to the last cell of the strip)
the selected particle would visit during its 
forward motion. The integer number $H$ will be called 
\emph{horizon depth}.

Either a red particle 
moves to one of the four neighboring 
cells with probabilities
$1-3r/4$ (down), 
$r/4$ (left), 
$r/4$ (right), 
and
$r/4$ (up)
if $H=0$,  or the horizon is empty, or the closest particle in the horizon is red. 
Otherwise, it 
moves to one of the four neighboring 
cells with probabilities 
$1-h$ (down), 
$h/2$ (left), 
$h/2$ (right), 
and
$0$ (up), see Figure~\ref{f:reticolo}.

Similarly, either 
a blue particle 
moves to one of the four neighboring 
cells with probabilities
$1-3r/4$ (up), 
$r/4$ (left), 
$r/4$ (right), 
and
$r/4$ (down)
if $H=0$,  or the horizon is empty, or the closest particle 
in the horizon is blue. 
Otherwise, the particle 
moves to one of the four neighboring 
cells with probabilities 
$1-h$ (up), 
$h/2$ (left), 
$h/2$ (right), 
and
$0$ (down).

The fact that the jumping probabilities change when an opposing particle 
is spotted inside the horizon will be addressed as 
\emph{anticipation mechanism}.
The parameter $r$ is called background noise. The model 
aims to describe two families of pedestrians one heading down and the other
heading up. This is precisely what happens in our model with  the 
red and blue particles if $r=0$. However, when $r$ is positive, 
different moves are allowed
in the system as it happens to real pedestrian crowds in motion --  sometimes 
they displace not following their prescribed best trajectory, but  
with random shifts due to external noise, such as sounds, light flashes, 
images, or obstacles.    
Note that if $r$ is small,  red and blue particles  experience 
an important downward and, respectively, an upward drift, which becomes 
smaller when $r$ increases and finally disappears at $r=1$, when 
the walk becomes perfectly symmetric. 

We introduced the parameter $r$ not only to mimic random real 
world shifts, but also for a technical reason. Indeed, in the case 
$H=0$, namely, when the anticipation effect is not considered, 
for $r=0$ our model would be completely trivial, indeed, red and blue 
particles would move one against the other and eventually would 
stop each others. A residual trivial 
motion will be present only in those columns 
populated by particles moving all in the same direction.  

In the following,  we  study the model for a wide choice of 
the parameters; the reader should always keep in mind that 
the values relevant for pedestrian flow scenarios are
$r\ll1$ and $h\gg0$. Indeed, a walker will change his direction 
of motion 
only in the presence of an opposing pedestrian (or other obstacle) and in such a case 
he will do it almost surely. 

The vertical boundaries are considered as occupied sites, that is 
to say, reflecting boundary conditions are imposed on those vertical 
boundaries of the strip $\Lambda$. On the other hand, the horizontal 
boundaries are considered filled with empty spots, 
so that a particle
on the first row trying to jump up will exit the system and, similarly,  
 a particle on the $L_2$--th row trying to jump down 
will exit. 
Particles which did exit the lattice, when selected for a move, 
will re--enter the strip with the same horizontal coordinate 
at row one if own color red, and at row $L_2$ if the particle is blue, provided the target site is empty. 
We have not considered strictly imposed vertical periodic boundary conditions -- 
the upper and the lower rows of the strip 
mimick the presence of doors at the end of the corridor (see, also, Appendix~\ref{s:bcp}).

\subsection{Observables}
\label{s:lane-obs}
Quantitative investigations of the model will be performed by means of the 
following observables. 
We  call \emph{upward current} at time 
$t$, the 
ratio between the total number of blue particles which exited the 
system through the top boundary and time. 
Similarly, 
we  call \emph{downward current} at time 
$t$, the 
ratio between the total number of red particles which exited the 
system through the bottom boundary and time. 
Note that both these currents are defined as positive numbers. 
The currents will be used to detect the presence of jamming in the 
system. More precisely, since the upward and the downward current will 
be approximatively equal in all the simulations, 
we will focus on the 
\emph{average current}, namely, the average between the upward and 
the downward currents. 

Additionally, to give a quantitative estimate of the presence of 
lanes in the system, we  define a suitable order parameter 
following closely  the ideas proposed in \cite{NS} and based on 
developments from  \cite{RL} done  in the framework of colloidal systems.

Fix the time $t$ and consider a particle labelled by $k\in\{1,\dots,N\}$
such that it lies in the lattice at time $t$.
 Let $n_{\rr{r},k}(t)$ the total number of red particles 
occupying cells belonging to the same column as the particle $k$. 
Furthermore, let $n_{\rr{b},k}(t)$ the total number of blue particles 
occupying cells belonging to the same column as the particle $k$. 
Then set 
\begin{equation}
\label{lane-op}
\phi(t)
=
\frac{1}{n(t)}\sum_{k=1}^{n(t)}
\Big[
     \frac{n_{\rr{r},k}(t)-n_{\rr{b},k}(t)}{n_{\rr{r},k}(t)+n_{\rr{b},k}(t)}{}
\Big]^2
\,.
\end{equation}
Note that in a state in which blue and red particles moved perfectly in 
separate lanes, 
the order parameter would be equal to one.
For disordered states, we expect $\phi(t)$ to be small, though 
strictly positive. 
In the next sections, we shall use the expression 
\textit{ordered configurations} 
when referring to configurations in which 
red and blue particle occupy different columns. 

\section{Numerical simulations}
\label{s:lane-sim}
We simulate the model introduced in Section~\ref{s:lane-mod} posed 
on the strip with side lengths
$L_1=50$ and 
$L_2=100$.
We fix a parameter $\rho$, called \emph{density}, and the total 
number of particles will then be $N=\rho L_1L_2$. The numbers of red 
and blue particles, $N_\text{r}$ and $N_\text{b}$, will differ at 
most by one and will be such that $N_\text{r}+N_\text{b}=N$.
The values of $H$, $h$, $r$, and $\rho$ will be specified 
 both in the forthcoming discussion of the numerical results and in the caption of the figures. 

All simulations are run for $8\times10^6$ time steps: remember that 
at each time step $N$ particles are randomly selected for motion. 
The order parameter is computed by averaging its value each $10^2$ 
time steps starting from the thermalization time $10^6$. 
The currents are computed by applying the definition (given in Section \ref{s:lane-obs})
at the end of the simulation.
The computed observables are 
very stable and the statistical errors are not significative, hence they are not 
reported in the pictures. 

Aiming to a good vizualization of the effects, the results will be presented by means of two different kind 
of graphs: configuration pictures and scatter plots. 
In configuration pictures, each point represents the position 
of a particle: red points stand for red particles and blue points  stand 
for blue particles. 
In scatter plots, either the current or the order parameter are 
reported for approximatively $20\times20$ different values of the 
considered parameters evenly spaced in the intervals specified 
in the graphs. No data interpolation is performed, each measured 
value corresponds to a square pixel in the picture.
The colors shown in scatter plots are 
adapted to picture data, but in all the pictures blue corresponds 
to the half of the maximum value in the plot. Moreover, for the values
below such half value we use gray tones, 
whereas for the values above it,  we use the following 
brilliant colors: magenta, 
red, orange, yellow, and green. 

\subsection{Preliminary simulations}
\label{s:prs}
As we already pointed out in Section \ref{s:lane-obs}, the order parameter $\phi$ is 
a positive number close to one for ordered configurations.
On the other hand, 
it is not clear how small such a parameter will be for
disordered configurations. Particularly,  we cannot infer that $\phi$ will be 
close to zero. Indeed, cf. \eqref{lane-op}, $\phi$ is defined 
as a sum of positive numbers, so that fluctuations in random 
configurations will add up and not cancel. 
Mainly for this reason, we perform a first study of the system 
for a wide choice of the parameter $r$. Obviously, we expect that 
for an $r$ not small,  the system will be essentially disordered. 
In this way, we will give a quantitative measure of the values 
that the order parameter $\phi$ should exhibit for disordered 
states.  

\begin{figure}[ht]
\begin{picture}(450,140)(5,0)
\put(0,0)
{
 \includegraphics[width=.4\textwidth,height=.4\textwidth]{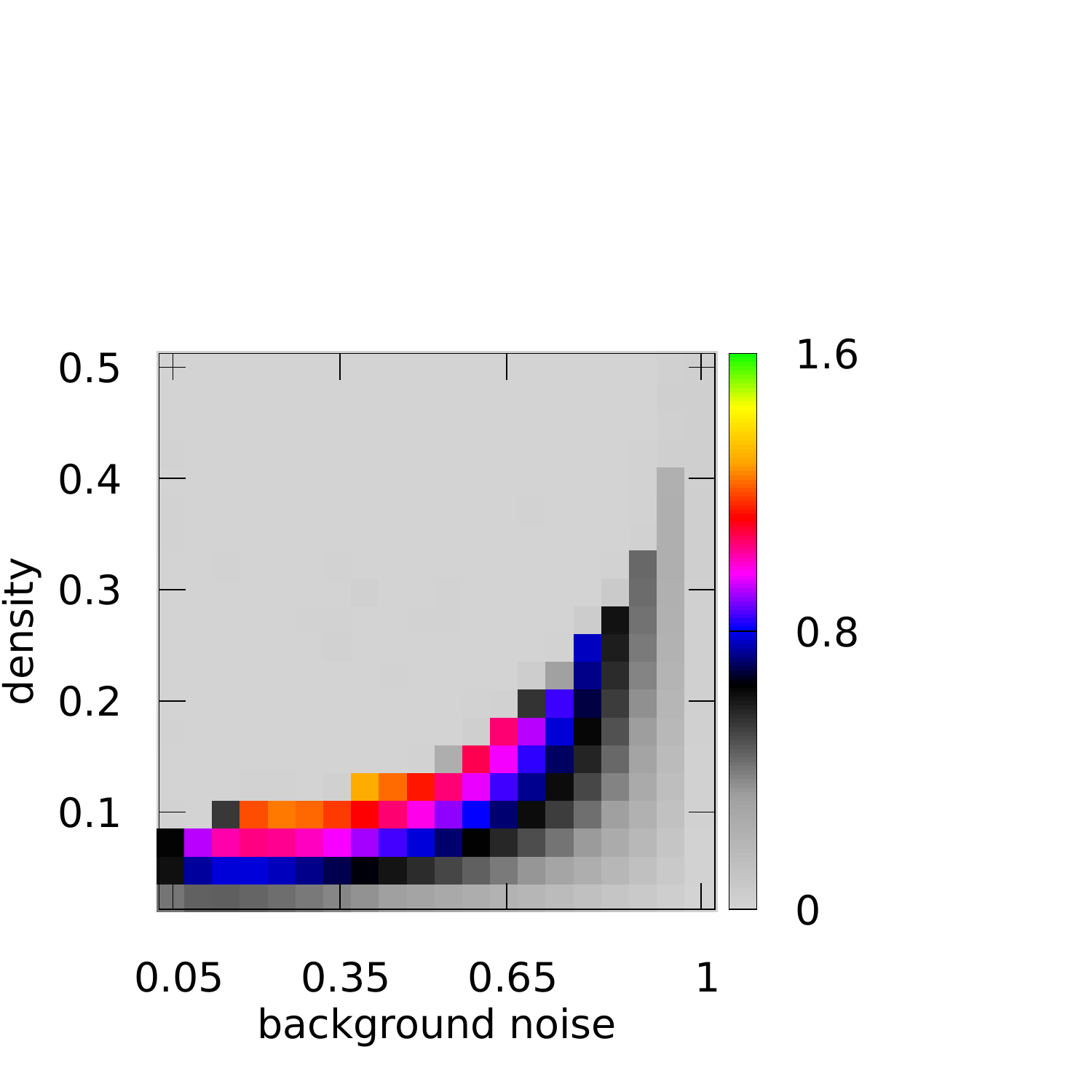}
}
\put(160,0)
{
 \includegraphics[width=.4\textwidth,height=.4\textwidth]{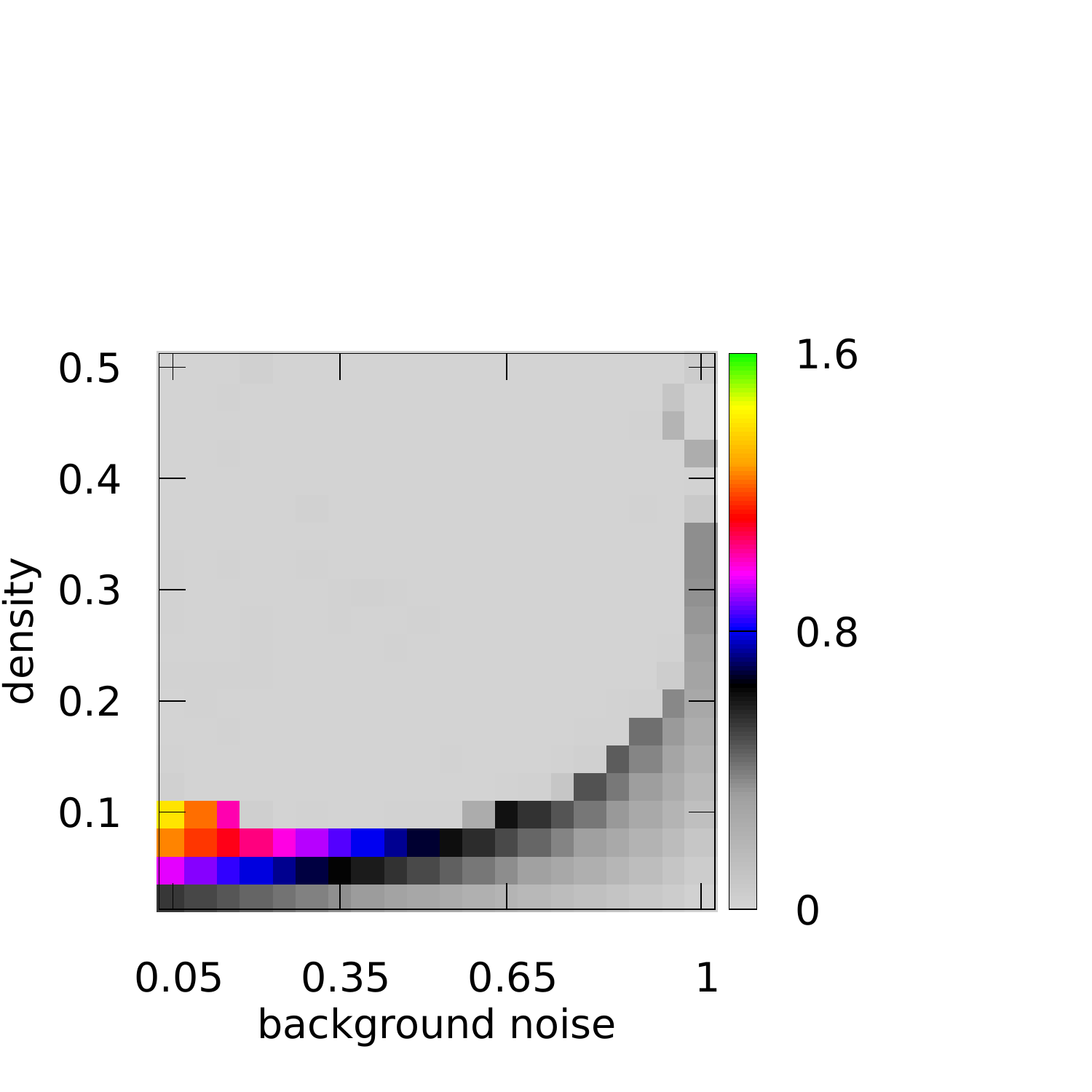}
}
\put(320,0)
{
 \includegraphics[width=.4\textwidth,height=.4\textwidth]{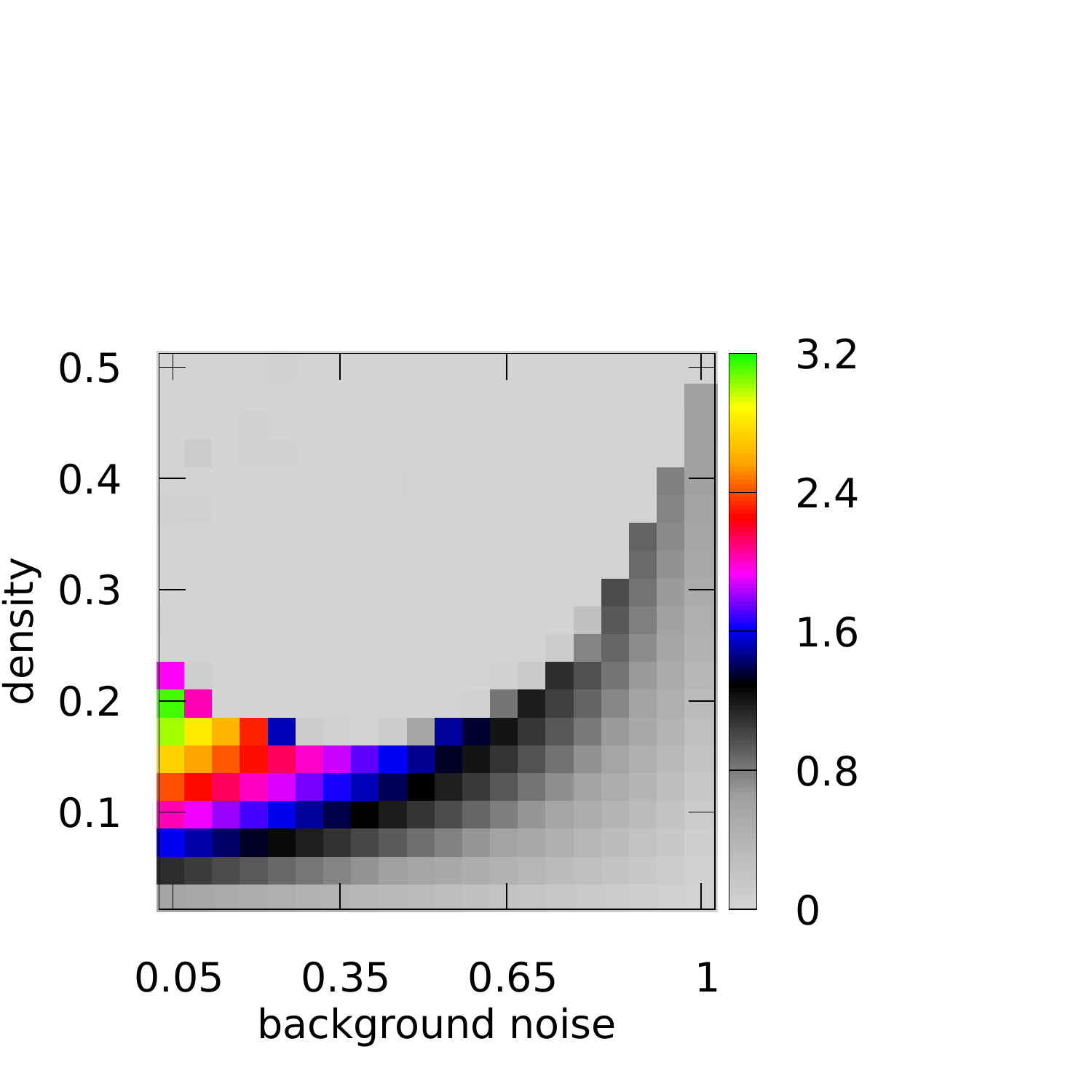}
}
\end{picture}
\caption{Scatter plot of the average current in the plane $r$--$\rho$
for 
$r\in[0.05,1]$,
$\rho\in[0.05,0.5]$,
$H=0$ (left), $H=5$ and $h=0.1$ (center), 
$H=5$ and $h=0.7$ (right).
}
\label{fig:n002} 
\end{figure}

In Figure~\ref{fig:n002},  we plot the average current in a scatter 
plot as a function of the background noise $r$ and the 
density $\rho$ for different choices of the horizon depth $H$ 
and of the lateral move probability $h$.  
As expected,  the highest values of the average current are 
found for low values of the background noise. Indeed, as already noted, when $r$ is small,  particles experience an important 
forward drift. 
Nevertheless, seeing the left diagram in the picture, we realize that when $H=0$ 
for some intermediate value of the density (focus, for instance, 
to the case $\rho=0.15$),  the current, which is negligible 
for $r$ small, becomes important if $r$ is mildly increased, 
but it eventually becomes zero for even larger values of 
the background noise. 
This effect is due to the fact that, for small values of the background noise,  
the dynamics is trapped in blocked (clogged) configurations, whereas a larger 
value of randomness in the dynamics helps particles to avoid blocking 
opponents restoring the global current to not zero values. 

\begin{figure}[ht]
\begin{picture}(450,140)(40,0)
\put(0,-70)
{
 \includegraphics[width=2.\textwidth,height=2.\textwidth]{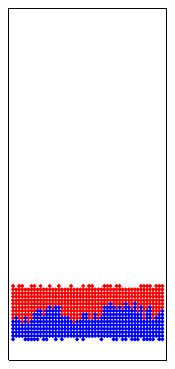}
}
\put(75,-72)
{
 \includegraphics[width=2.\textwidth,height=2.\textwidth]{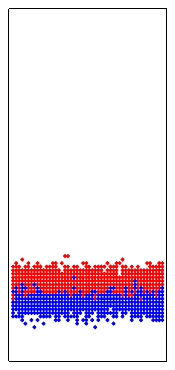}
}
\put(150,-72)
{
 \includegraphics[width=2.\textwidth,height=2.\textwidth]{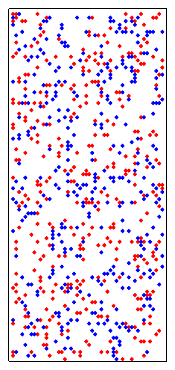}
}
\put(225,-68.75)
{
 \includegraphics[width=2.\textwidth,height=2.\textwidth]{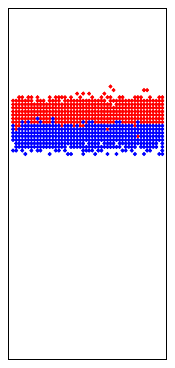}
}
\put(300,-72)
{
 \includegraphics[width=2.\textwidth,height=2.\textwidth]{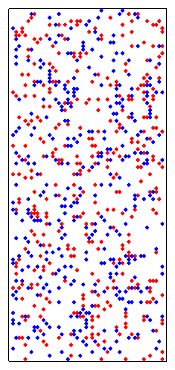}
}
\put(375,-72)
{
 \includegraphics[width=2.\textwidth,height=2.\textwidth]{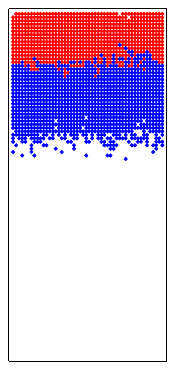}
}
\end{picture}
\caption{From the left to the right 
it is depicted 
the final configuration of the simulations for the 
cases 
$H=0$, $r=0.1$, and $\rho=0.15$ (first graph), 
$H=0$, $r=0.5$, and $\rho=0.15$ (second graph), 
$H=0$, $r=0.7$, and $\rho=0.15$ (third graph), 
$H=5$, $h=0.1$, $r=0.5$, and $\rho=0.15$ (fourth graph), 
$H=5$, $h=0.7$, $r=0.5$, and $\rho=0.15$ (fifth graph),
and
$H=5$, $h=0.7$, $r=0.8$, and $\rho=0.45$ (sixth graph).
}
\label{fig:n003} 
\end{figure} 

This is illustrated in Figure~\ref{fig:n003} where 
it is reported 
the final configuration 
of the system in the simulations for 
the values of the parameters specified in the caption. 
The first three panels on the left show that considering 
larger values of $r$, in absence of the anticipation mechanism, 
allows to avoid blocking configurations. In this respect,  
randomness favors transport.
It is interesting to remark that a similar phenomenon was 
found by the authors in \cite[Figures~6.14 and 6.15]{CKMSS16}, where 
it was remarked that the so called residence time it is a not monotonic 
function of the lateral displacement probability. We recall 
that the residence 
time was defined in {\em loc. cit.} as the typical time that a particle 
started at one side of the strip needs to cross the whole strip and exit from 
the opposite boundary. Consequently, the residence time 
and the current are closely related quantities. 

Another interesting phenomenon can be observed  comparing the second, the fourth and the fifth panels in Figure~\ref{fig:n003}.
In these three cases, the values of background noise and density 
are not changed, but the anticipation effect is introduced 
and the lateral move probability is changed. 
The pictures show that adding the anticipation mechanism with a 
sufficiently large lateral move probability blocking 
configurations can be avoided. 
The fact that anticipation helps transport in a wide region 
of the parameter space is also evident from 
the current graphs in Figure~\ref{fig:n002}, but the configurations 
reported in Figure~\ref{fig:n003} provide a striking evidence.
The sixth configuration in Figure~\ref{fig:n003} shows that for very 
large values of the density, even a large lateral move probability is 
not sufficient to the restore current, and consequently, the dynamics is eventually 
trapped in a blocked configuration. 

\begin{figure}[ht]
\begin{picture}(450,140)(5,0)
\put(0,0)
{
 \includegraphics[width=.4\textwidth,height=.4\textwidth]{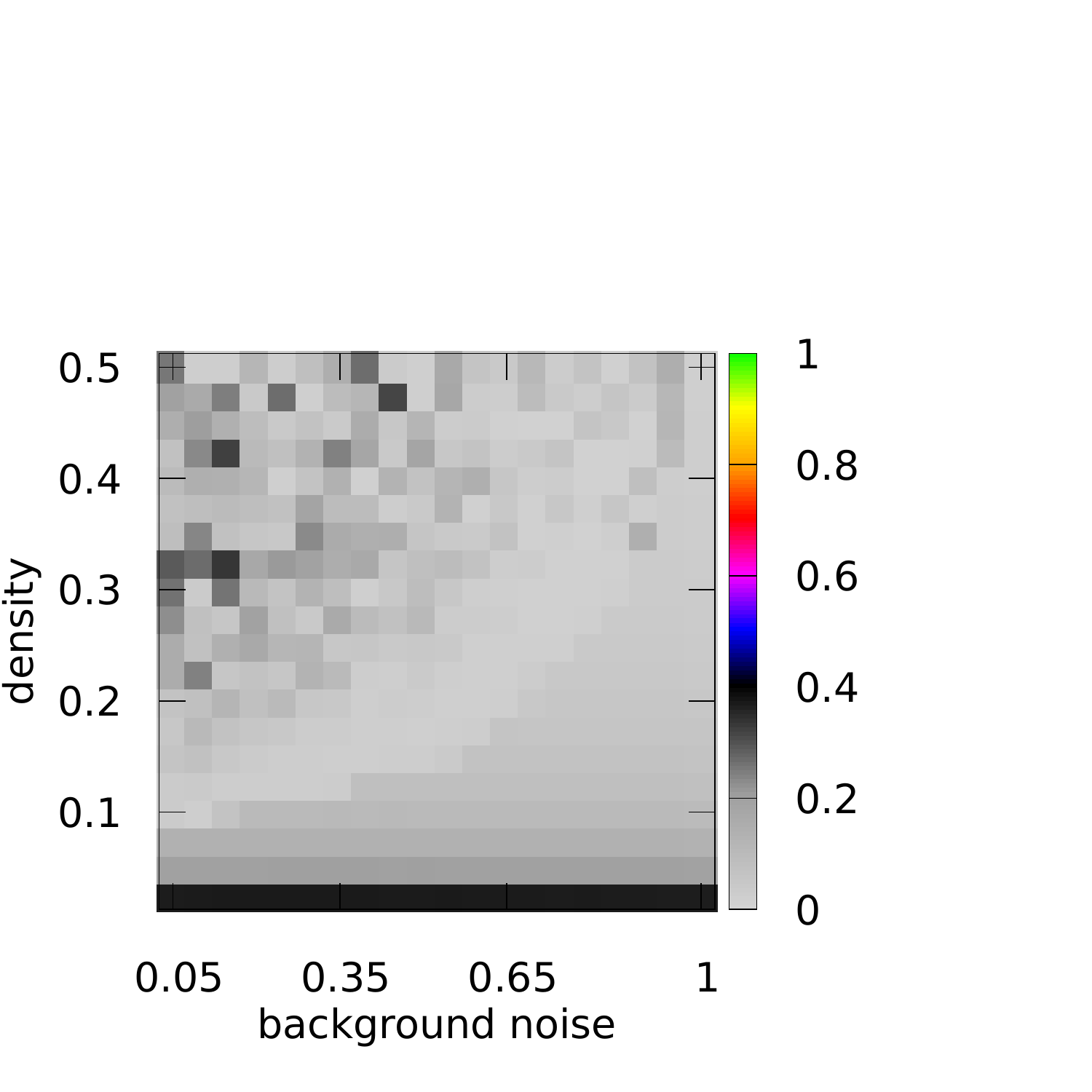}
}
\put(160,0)
{
 \includegraphics[width=.4\textwidth,height=.4\textwidth]{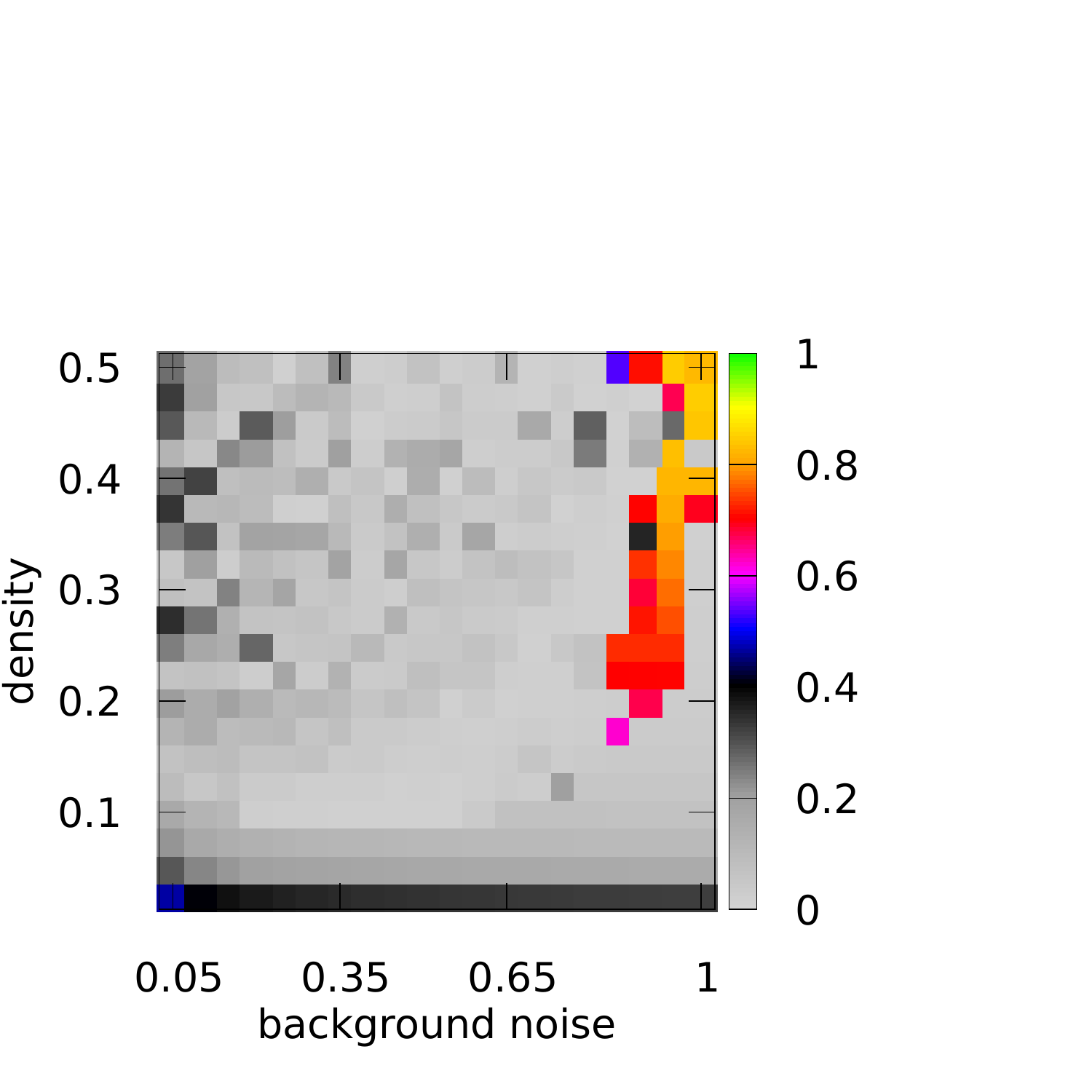}
}
\put(320,0)
{
 \includegraphics[width=.4\textwidth,height=.4\textwidth]{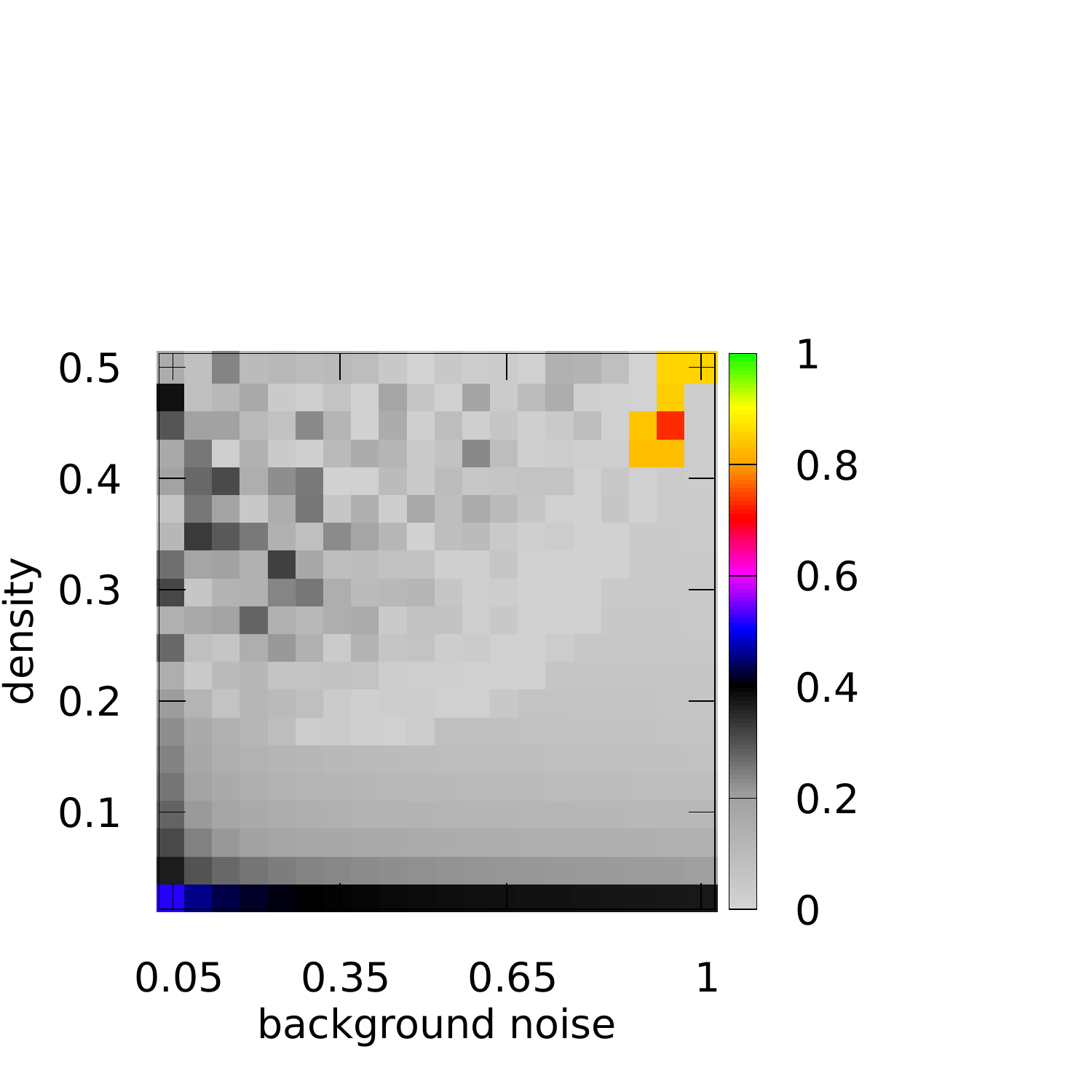}
}
\end{picture}
\caption{Scatter plot of the order parameter $\phi$ 
in the plane $r$--$\rho$ for the same case considered in Figure~\ref{fig:n002}.
}
\label{fig:n004} 
\end{figure} 

In Figure~\ref{fig:n004}, we finally come to the main target of this 
section, namely, the graph of the order parameter in the plane 
$r$--$\rho$. The scatter plot is reported for the 
same cases considered in Figure~\ref{fig:n002}. 
In all the cases, we do not expect lane formation, due to the 
rather high values of the background noise considered in the 
pictures. Indeed, the graphs in Figure~\ref{fig:n004} show 
that the order parameter is approximatively constant in the 
whole region considered in the simulations. Moreover, 
the stationary value of the order parameter ranges 
between $0.1$ and $0.3$. Hence, in the sequel of our 
discussion we will consider such a value as the reference 
point of the order parameter for completely 
disordered configurations.  
The small islands in the central and right panel 
corresponding to higher values of the order parameter 
can be neglected since they are observed in correspondence of
blocked configurations.
The relative high value of $\phi$ 
is just a random value depending on the random  initial condition, 
for instance in the case illustrated in the 
sixth panel in Figure~\ref{fig:n003}.  This is due to the fact that, 
in the final configuration,  many red particles remained blocked 
outside the lattice, hence in each column   a majority 
of red particles is present. This yields  in a rather high value of the 
order parameter. 

\subsection{Small background noise}
\label{s:sbn}
We now focus on the most interesting part of the parameter space, 
namely, the region with small background noise. 

\begin{figure}[ht]
\begin{picture}(450,110)(5,0)
\put(0,0)
{
 \includegraphics[width=.24\textwidth,height=.24\textwidth]{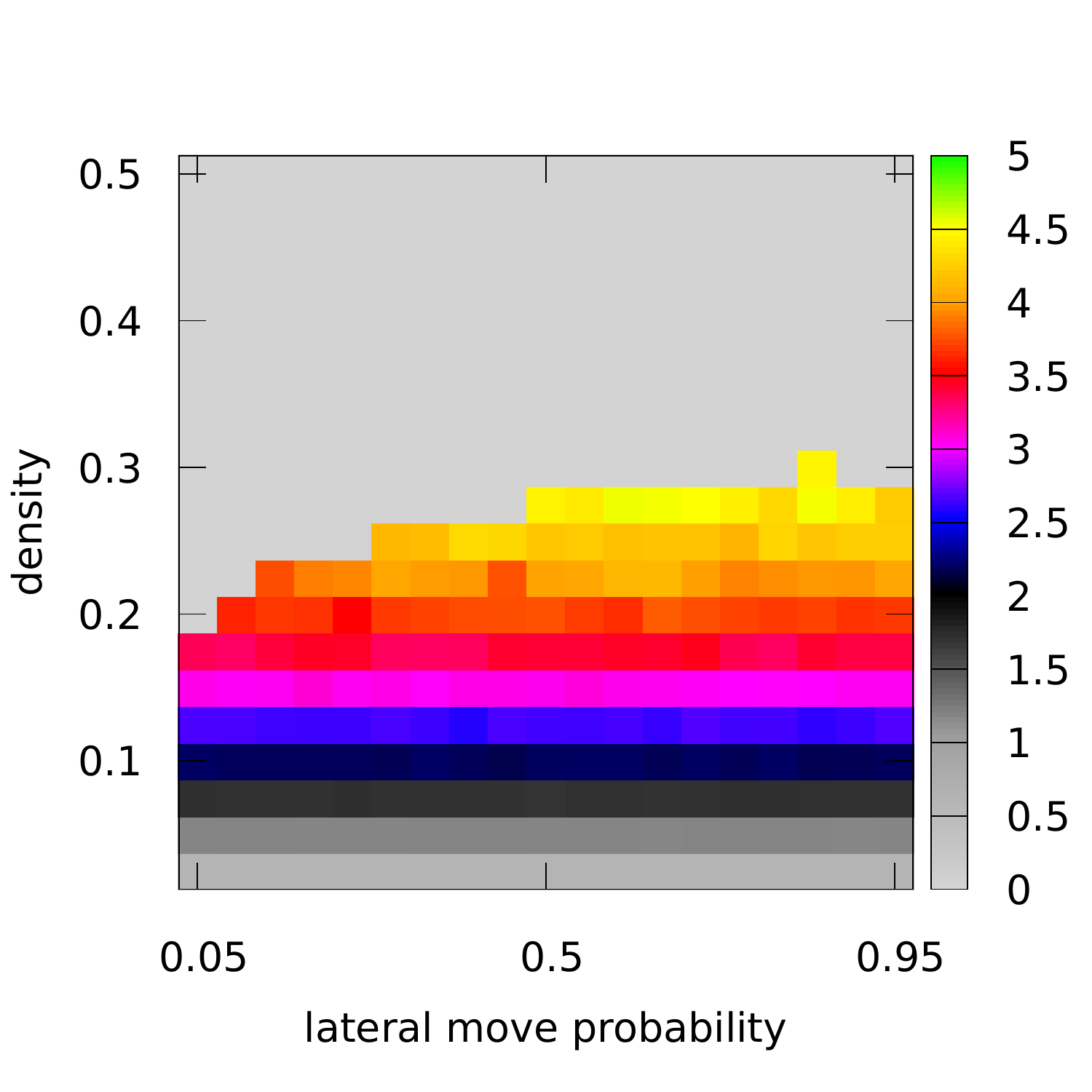}
}
\put(120,0)
{
 \includegraphics[width=.24\textwidth,height=.24\textwidth]{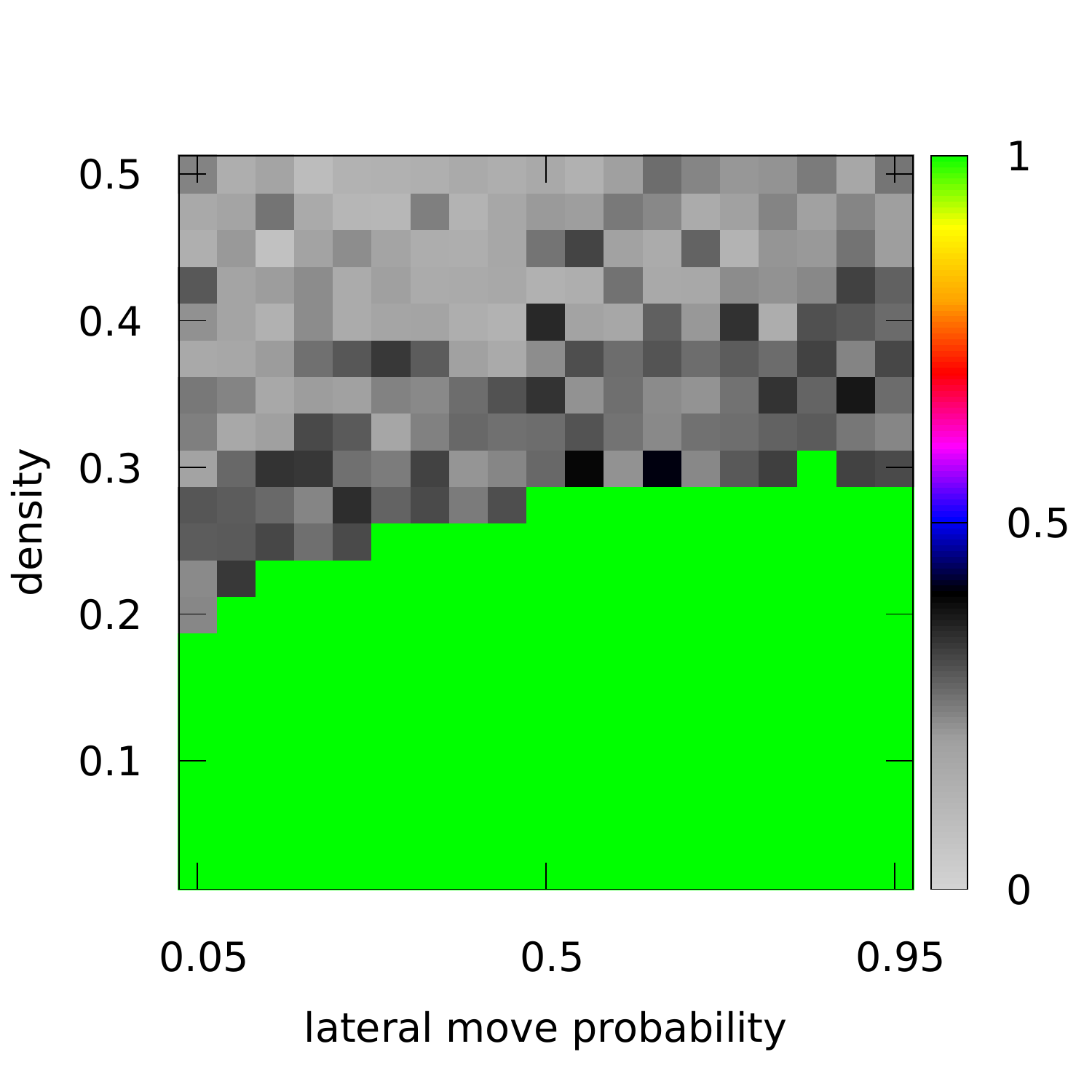}
}
\put(240,0)
{
 \includegraphics[width=.24\textwidth,height=.24\textwidth]{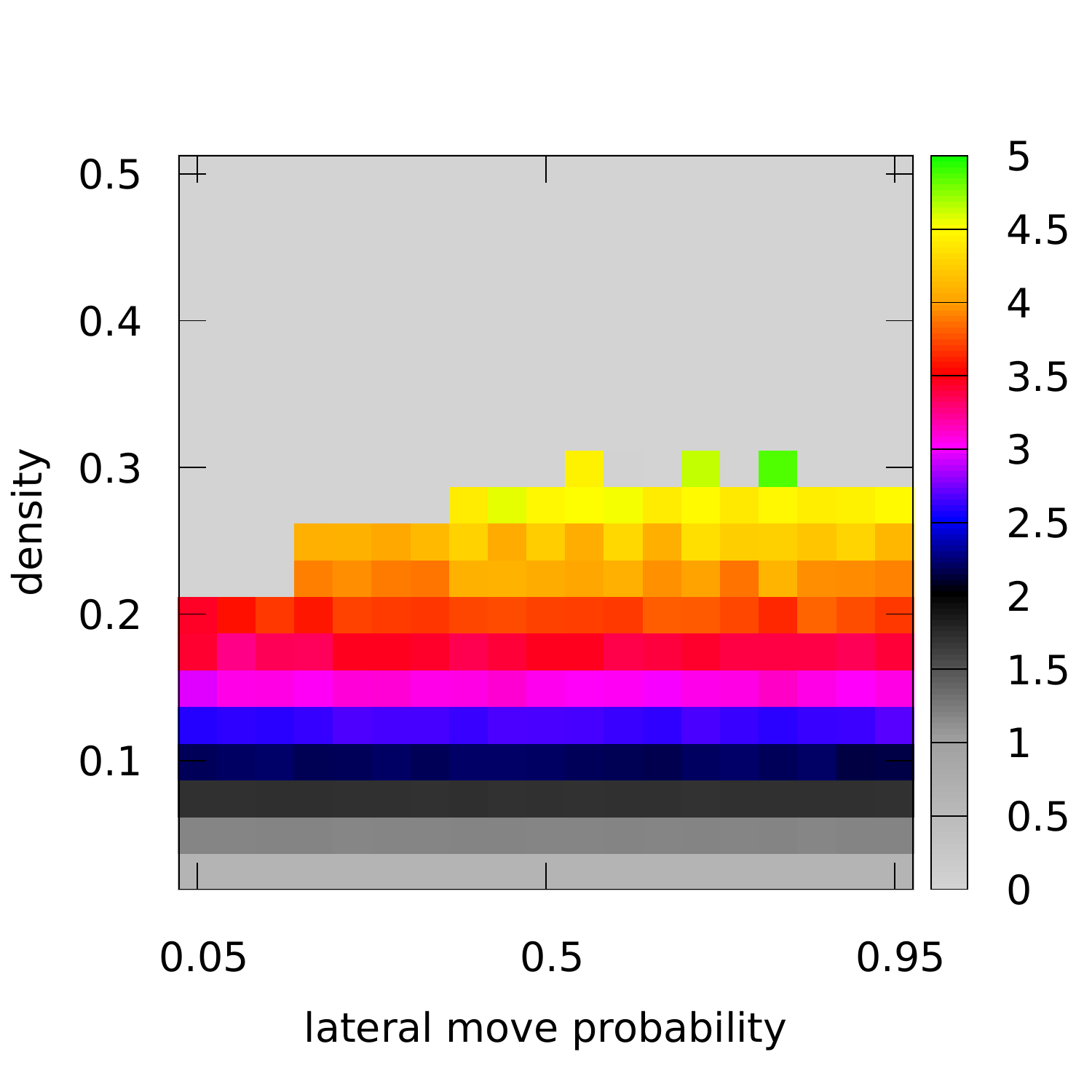}
}
\put(360,0)
{
 \includegraphics[width=.24\textwidth,height=.24\textwidth]{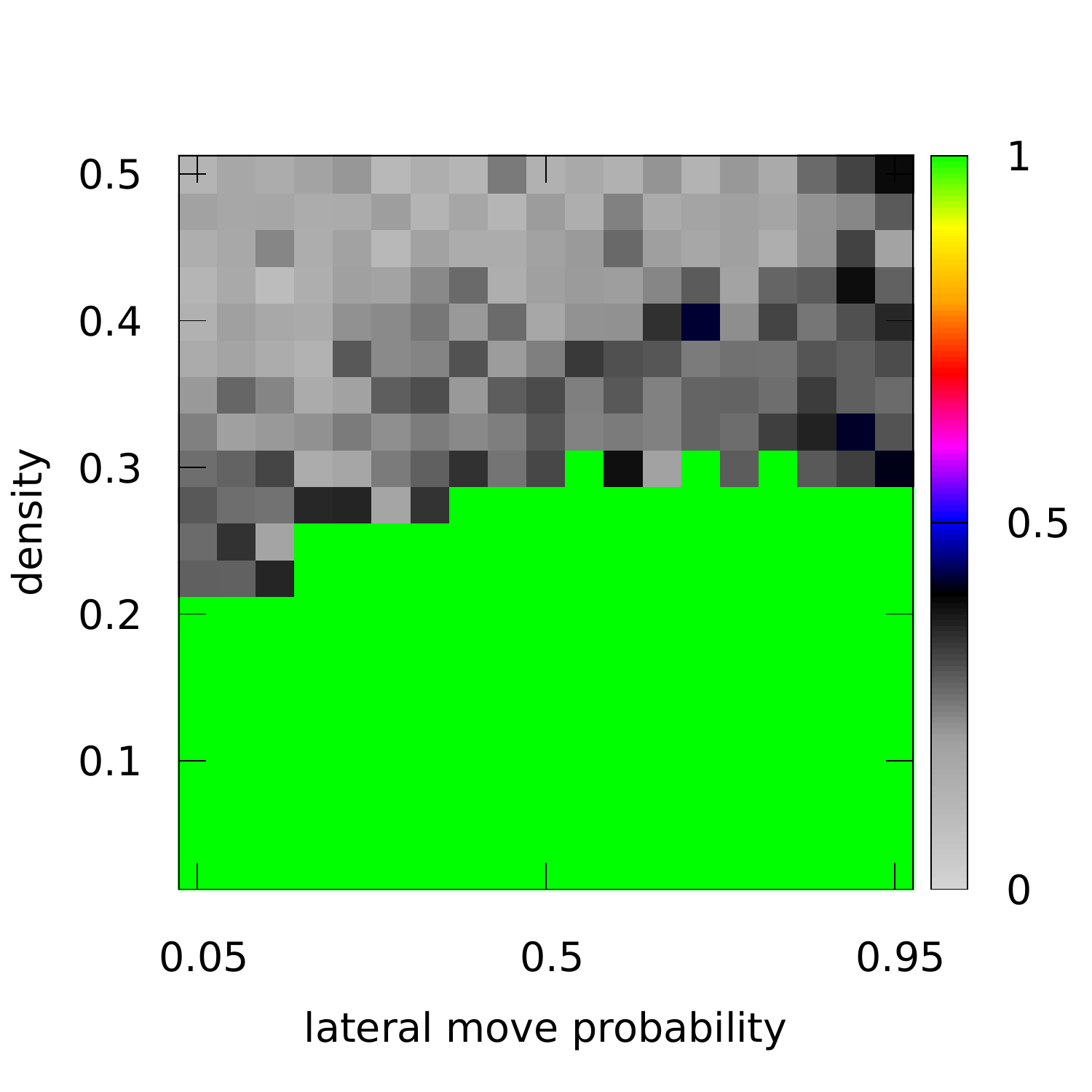}
}
\end{picture}
\caption{Scatter plot of the current (first and third panel) 
and order parameter $\phi$ (second and fourth panel)
in the plane $h$--$\rho$ for zero background noise,
$h\in[0.05,0.95]$,
$\rho\in[0.05,0.5]$,
$H=5$ (first two panels), 
and 
$H=20$ (third and fourth panel).
}
\label{fig:n005} 
\end{figure}

In Figure~\ref{fig:n005}, we have reported the results of our 
simulations at $r=0$, namely, when no background noise is present
so that the sole mechanism present in the dynamics is anticipation:
particles move in their prescribed forward direction unless 
an opposing particle is spotted inside the horizon region. 
In the picture, we show scatter plots in the 
plane $h$--$\rho$ for both the average current 
and the order parameter $\phi$. 

The left panel gives evidence that at any value of the lateral move 
probability, the average current increases with the density if 
this is sufficiently small, i.e., smaller that about $0.3$. 
On the other hand, when
such a value is reached, the dynamics freezes in blocked configurations, 
and hence, the currents suddenly drop to zero. 
It is quite remarkable that {\em the current, when different from zero, does 
not depend very much on the lateral move probability $h$.} 
On the other hand, as we have already noted in the 
above Section~\ref{s:prs}, for intermediate values of the density 
the anticipation mechanism helps transport, in the sense that 
the freezing of the dynamics occurs at larger value of the 
density if $h$ is large. 

\begin{figure}[ht]
\begin{picture}(450,140)(40,0)
\put(0,-72)
{
 \includegraphics[width=2.\textwidth,height=2.\textwidth]{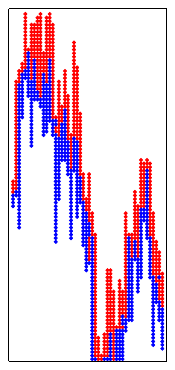}
}
\put(75,-72)
{
 \includegraphics[width=2.\textwidth,height=2.\textwidth]{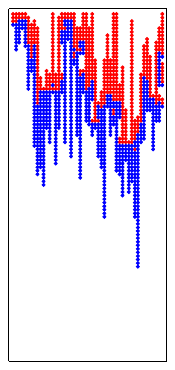}
}
\put(150,-72)
{
 \includegraphics[width=2.\textwidth,height=2.\textwidth]{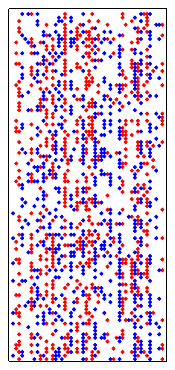}
}
\put(225,-72)
{
 \includegraphics[width=2.\textwidth,height=2.\textwidth]{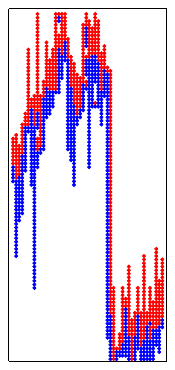}
}
\put(300,-72)
{
 \includegraphics[width=2.\textwidth,height=2.\textwidth]{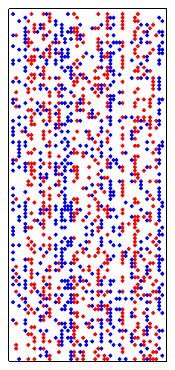}
}
\put(375,-72)
{
 \includegraphics[width=2.\textwidth,height=2.\textwidth]{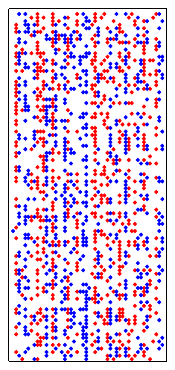}
}
\end{picture}
\caption{From the left to the right 
it is depicted 
the final configuration of the simulations for the 
cases 
$r=0$, $\rho=0.275$, 
$H=5$  and $h=0.05$ (first graph), 
$H=5$, and $h=0.45$ (second graph), 
$H=5$, and $h=0.50$ (third graph), 
$H=20$ and $h=0.05$ (fourth graph), 
$H=20$ and $h=0.45$ (fifth graph),
and
$H=20$ and $h=0.50$ (sixth graph).
}
\label{fig:n006} 
\end{figure}

To emphasize this point aspect in a better way, we have plotted in 
Figure~\ref{fig:n006} the final simulation configuration 
of the system at density $\rho=0.275$,  for different values of the 
lateral move probability $h$. The pictures shows that if $h$ is 
small the dynamics is eventually trapped in a blocked configuration 
whereas as $h$ is increased no freezing is observed, at least on the 
time scale we considered, and the current results to be different 
from zero. 

We finally remark that, for $r=0$, if the dynamics is not trapped 
then the 
order is perfect, namely, $\phi=1$ is reached. In other words,  {\em in the absence 
of the background noise, the anticipation mechanism 
guarantees a perfect lane formation, provided 
the dynamics is not frozen in blocked configurations}. In our opinion, this is a very valuable result, since it states that lanes forming in counterfows can 
be explained just as a consequence of the anticipation mechanism.

Data referring to the case $H=20$ and reported in Figures~\ref{fig:n005} 
and \ref{fig:n006} can be discussed similarly;  the only difference
is that the effect of the anticipation mechanism is slightly stronger.
This fact can be observed both in Figure~\ref{fig:n005} and \ref{fig:n006}.

We  test if the anticipation mechanism is robust 
with respect to the background noise, that is to say, 
if its ability to form lanes is still valid for $r$ 
different from zero.

\begin{figure}[ht]
\begin{picture}(450,140)(5,0)
\put(0,0)
{
 \includegraphics[width=.4\textwidth,height=.4\textwidth]{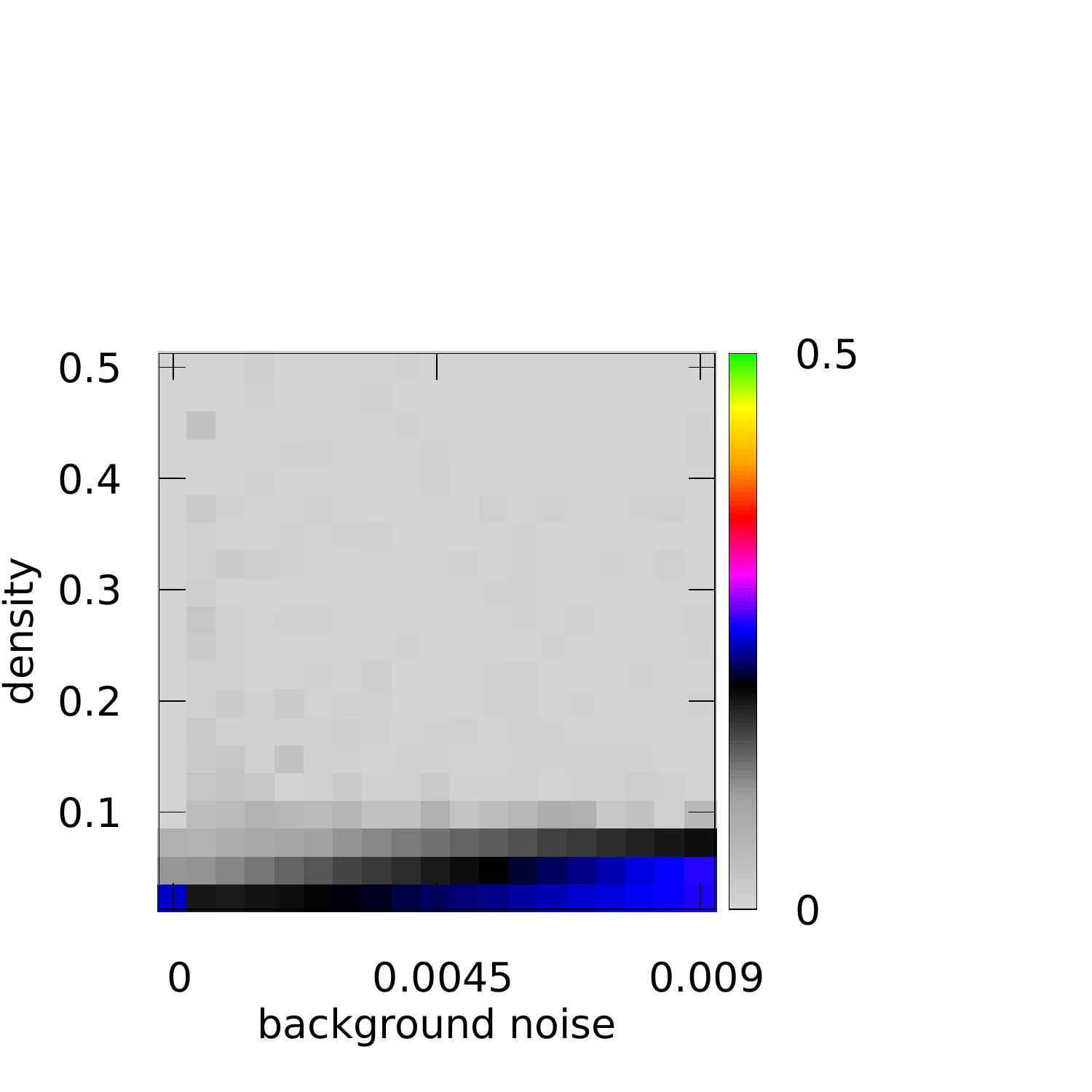}
}
\put(160,0)
{
 \includegraphics[width=.4\textwidth,height=.4\textwidth]{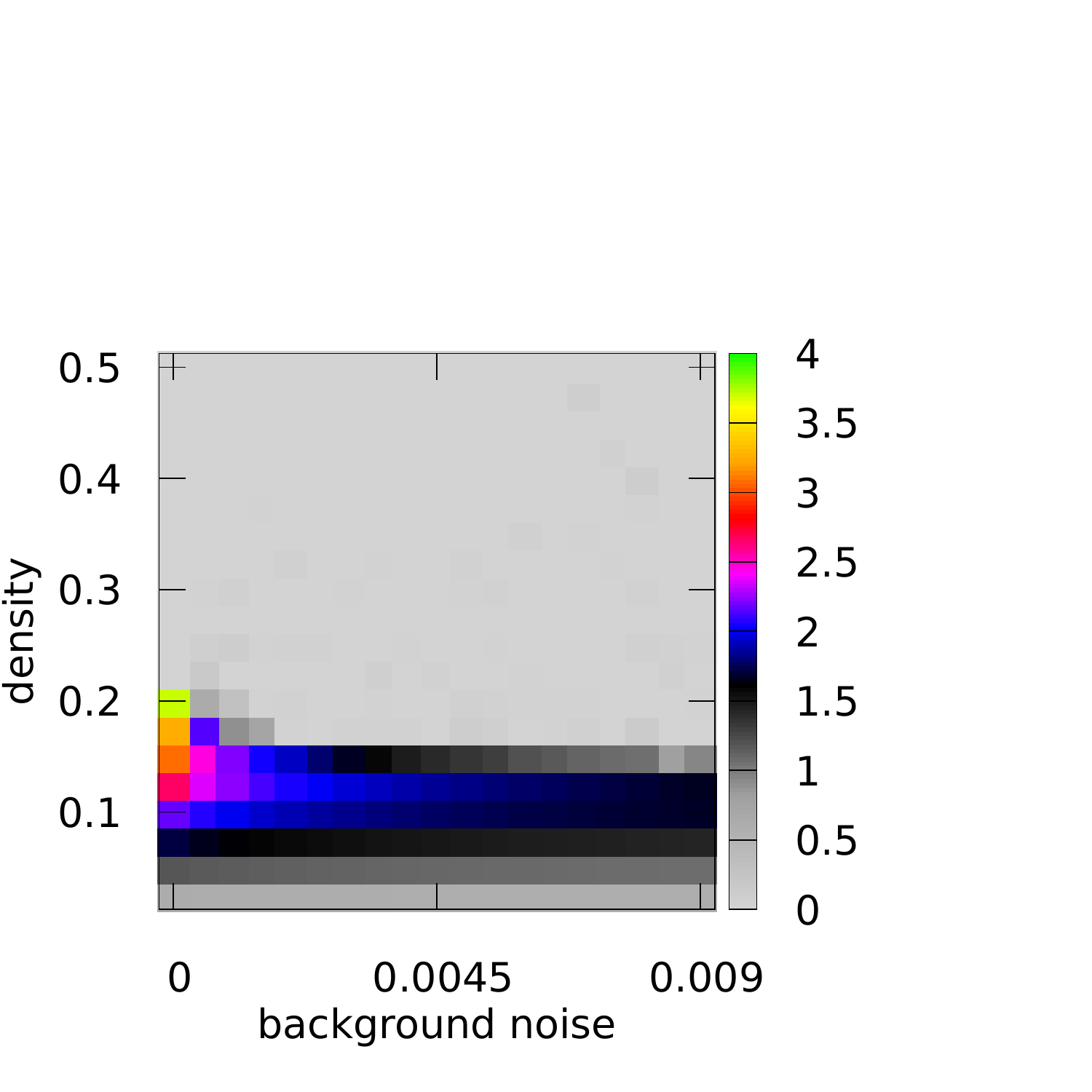}
}
\put(320,0)
{
 \includegraphics[width=.4\textwidth,height=.4\textwidth]{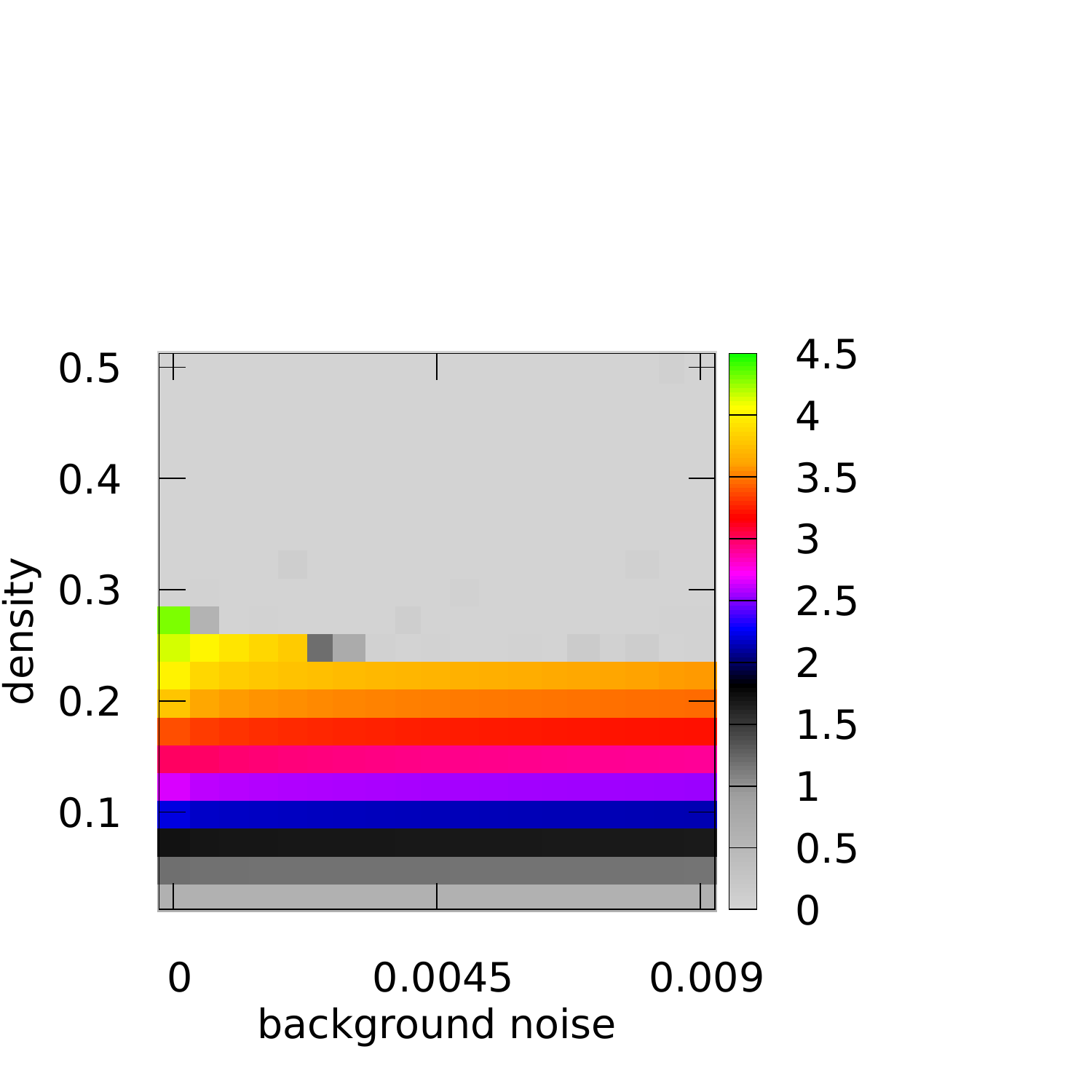}
}
\end{picture}
\caption{Scatter plot of the average current in the plane $r$--$\rho$
for 
$r\in[0,0.09]$,
$\rho\in[0.05,0.5]$,
$H=0$ (left), $H=5$ and $h=0.11$ (center), 
$H=5$ and $h=0.71$ (right).
}
\label{fig:n007} 
\end{figure} 

\begin{figure}[ht]
\begin{picture}(450,140)(5,0)
\put(0,0)
{
 \includegraphics[width=.4\textwidth,height=.4\textwidth]{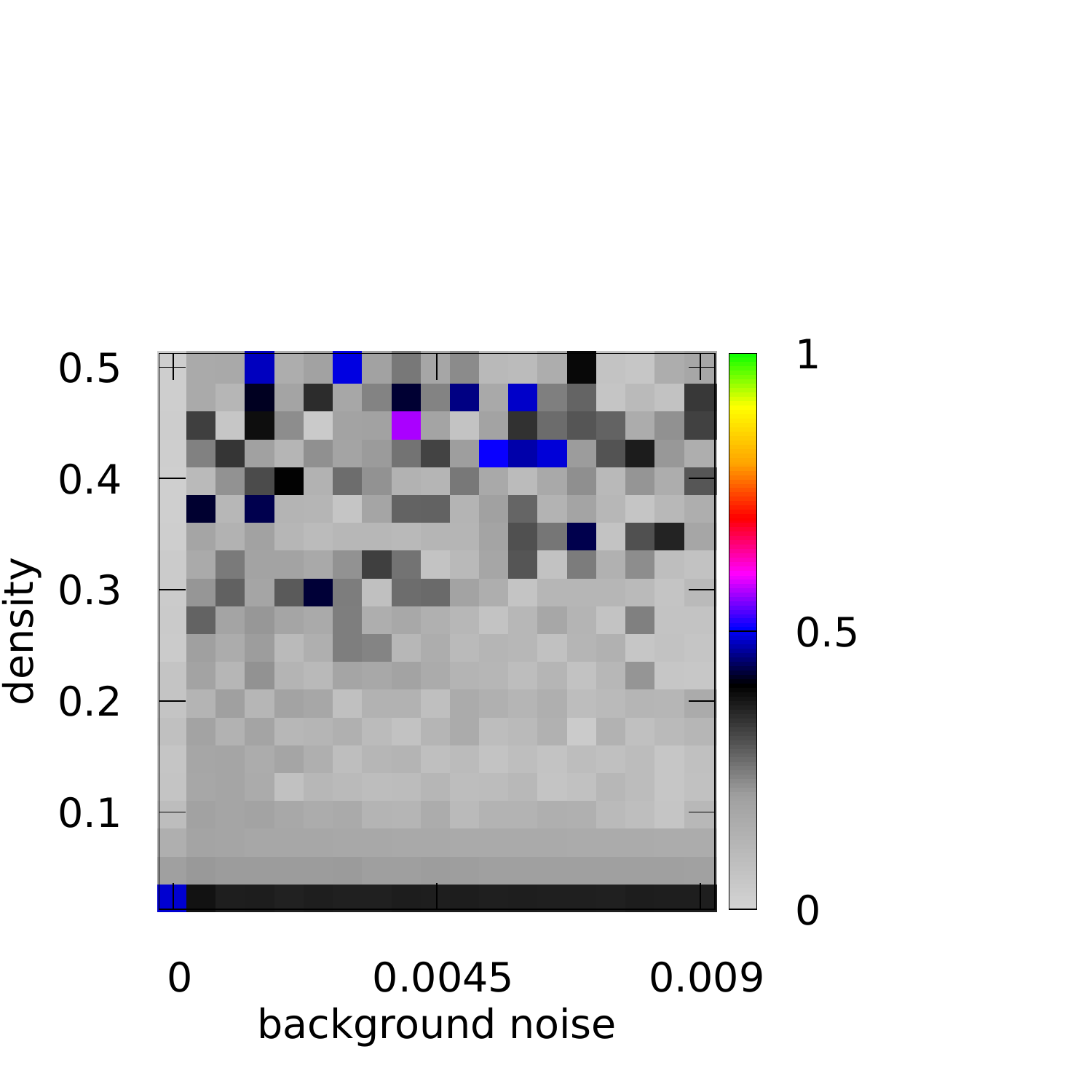}
}
\put(160,0)
{
 \includegraphics[width=.4\textwidth,height=.4\textwidth]{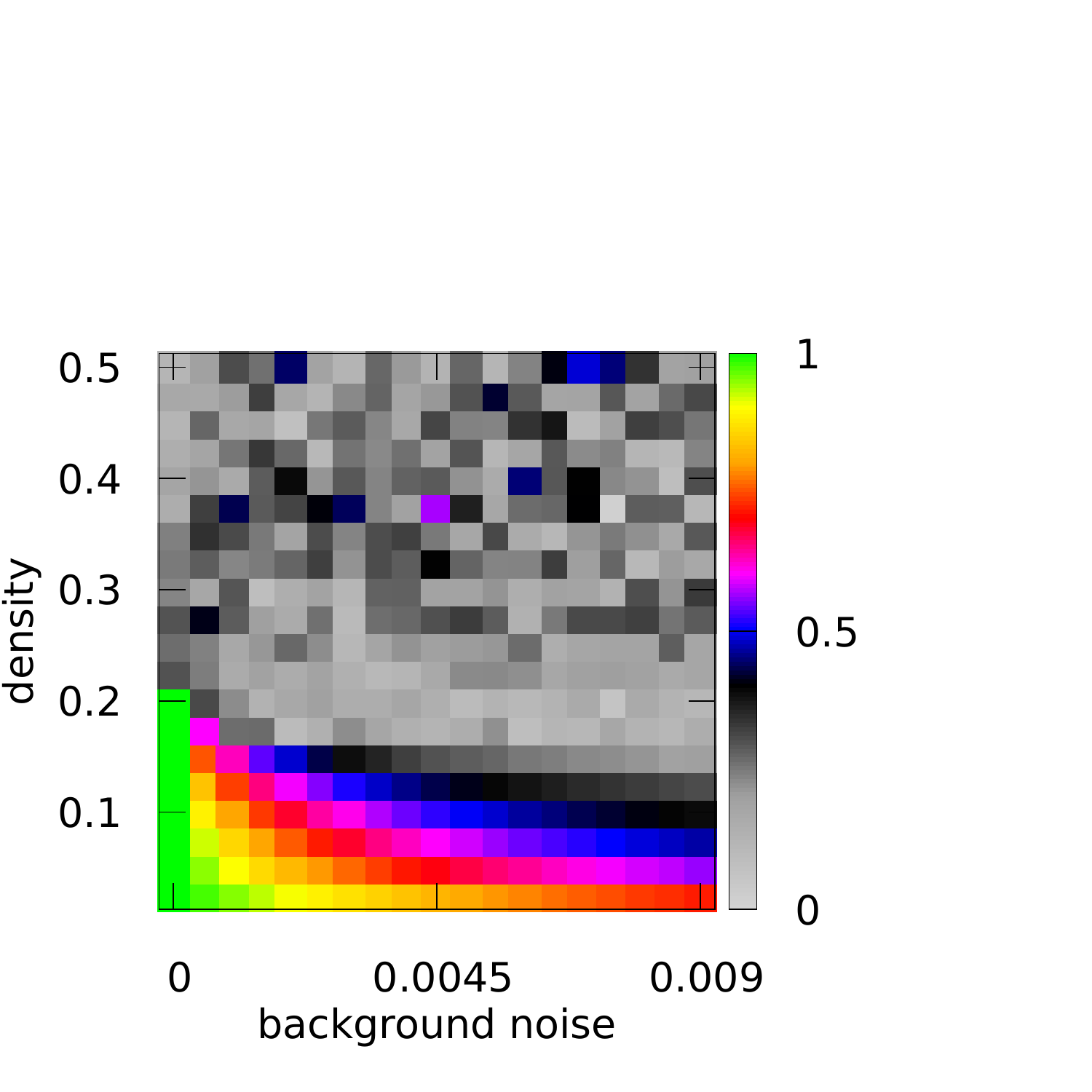}
}
\put(320,0)
{
 \includegraphics[width=.4\textwidth,height=.4\textwidth]{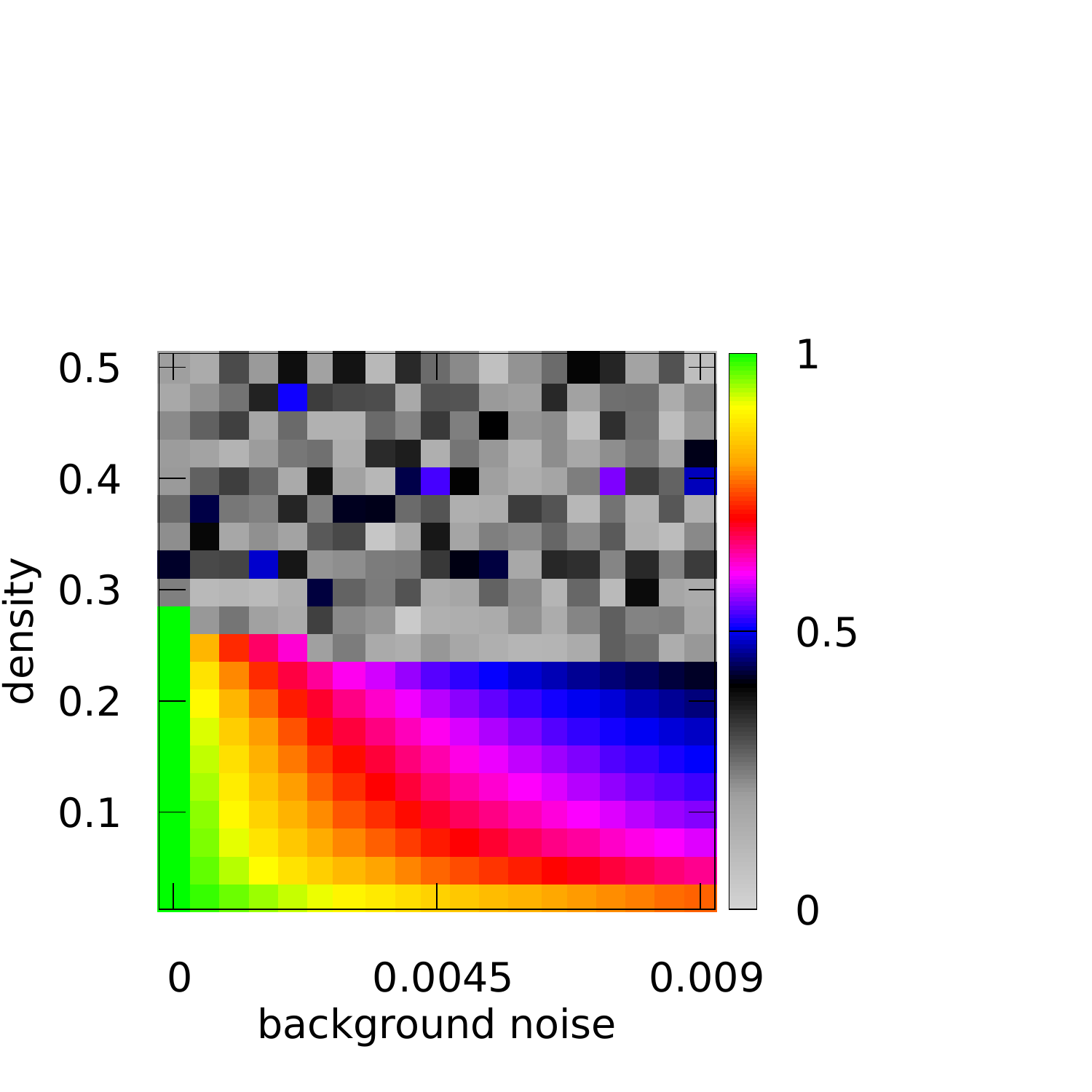}
}
\end{picture}
\caption{Scatter plot of the order parameter $\phi$ 
in the plane $r$--$\rho$ for the same case considered in Figure~\ref{fig:n007}.
}
\label{fig:n008} 
\end{figure} 

In Figures~\ref{fig:n007} and \ref{fig:n008} we report 
the scatter plot of the average current and the order parameter 
on the $r$--$\rho$ plane for different values of the lateral 
move probability. In particular, the left panel refers to the case in which 
the anticipation mechanism is not present in the dynamics. 

We pinpoint here a behavior which is very similar to the one discussed in the 
case $r=0$. In particular, we notice that, for a fixed value 
of the density $\rho$, increasing the 
lateral move probability avoids the freezing of the dynamics 
at larger values of the density $\rho$. Note also, 
that the current does not depend very much on the value of the disorder 
parameter in the considered range. This is quite obvious 
since in these pictures we are focussing on a very tiny slice 
of the part of the 
graphs shown in Figures~\ref{fig:n002} and \ref{fig:n004},  very
close to the vertical axis. 

Finally, we remark that Figure~\ref{fig:n008} shows that the anticipation 
mechanism is able to explain lane formation also in the presence of a 
weak background noise. Such an order is eventually destroyed 
if the parameter $r$ becomes too large as underligned by the data 
reported in Figure~\ref{fig:n004}.


\section{Discussion}
\label{s:dis}
As closing note, in the same line of thinking as in Ref. \cite{Sum06},  we argue that the key towards an even deeper understanding of  a collective behavior like lane formation lies in identifying the principles of the
behavioral algorithms followed by each individual,   and also, in answering the question: How does information flow among the pedestrians? Addressing this question requires the embedding in our model of fine environmental psychology information as well as aspects of the psychology of groups. We have not touched these aspects at all in this contribution. This can be seen as further work. On the hand,  for the  presented bi--directional pedestrian flow scenario, given two population sizes walking within  the strip $\Lambda$,  we are convinced that the simple combination of just  $3$ parameters  is sufficient to predict the formation of lanes. These parameters are the horizon depth $H$, the lateral move probability $h$, and the background noise $r$. This level of complexity is much lower than what usually the social force model is offering. Furthermore, our three parameters have a clear physical meaning. The harder to identify is eventually the background noise  level $r$, which on top of everything is also prone to different modeling interpretations and incorporates very much the specifics of the local conditions (geometry of the building, local traffic, etc.).

Two main results stand out:
\begin{itemize}
\item[(A)] In the absence 
of the background noise, the anticipation mechanism 
guarantees a perfect lane formation, provided 
the dynamics is not frozen in blocked configurations;
\item[(B)]  The current does 
not depend very much on the lateral move probability $h$.
\end{itemize}

An interesting question is to which extent (A) and (B)  hold 
 if pedestrians would be  moving at different speeds?
This question, connecting pedestrian flow to traffic flow matters,  could be addressed rather naturally using a continuous time 
version of the present model in which pedestrian moving at different 
speeds would be modelled by particles moving with different rates. 

\begin{acknowledgments} We thank Prof. Rutger van Santen (Eindhoven, NL) for very fruitful discussions on closely related matters.
ENMC thanks the \'ENS de Paris for the very kind hospitality in the 
period in which part of this work has been done. 
\end{acknowledgments}

\appendix
\section{Strict periodic boundary conditions}
\label{s:bcp}
As we have already mentioned in the above discussion, we considered 
"not strictly imposed vertical periodic boundary conditions". 
Our choice is motivated by the fact that the upper and the lower 
boundaries are thought as two open doors for the pedestrian motion. 
In this appendix, we show some minimal results obtained in the case when vertical periodic 
boudary conditions are strictly imposed as it is usually the case in the 
statistical mechanics of lattice models. By ``strictly imposed"
we simply mean that the 
updating rule is covariant in the lattice and the rows $0$ and $L_2+1$ 
are respectively identified with the rows $L_2$ and $1$. 

\begin{figure}[ht]
\begin{picture}(450,110)(5,0)
\put(0,0)
{
 \includegraphics[width=.24\textwidth,height=.24\textwidth]{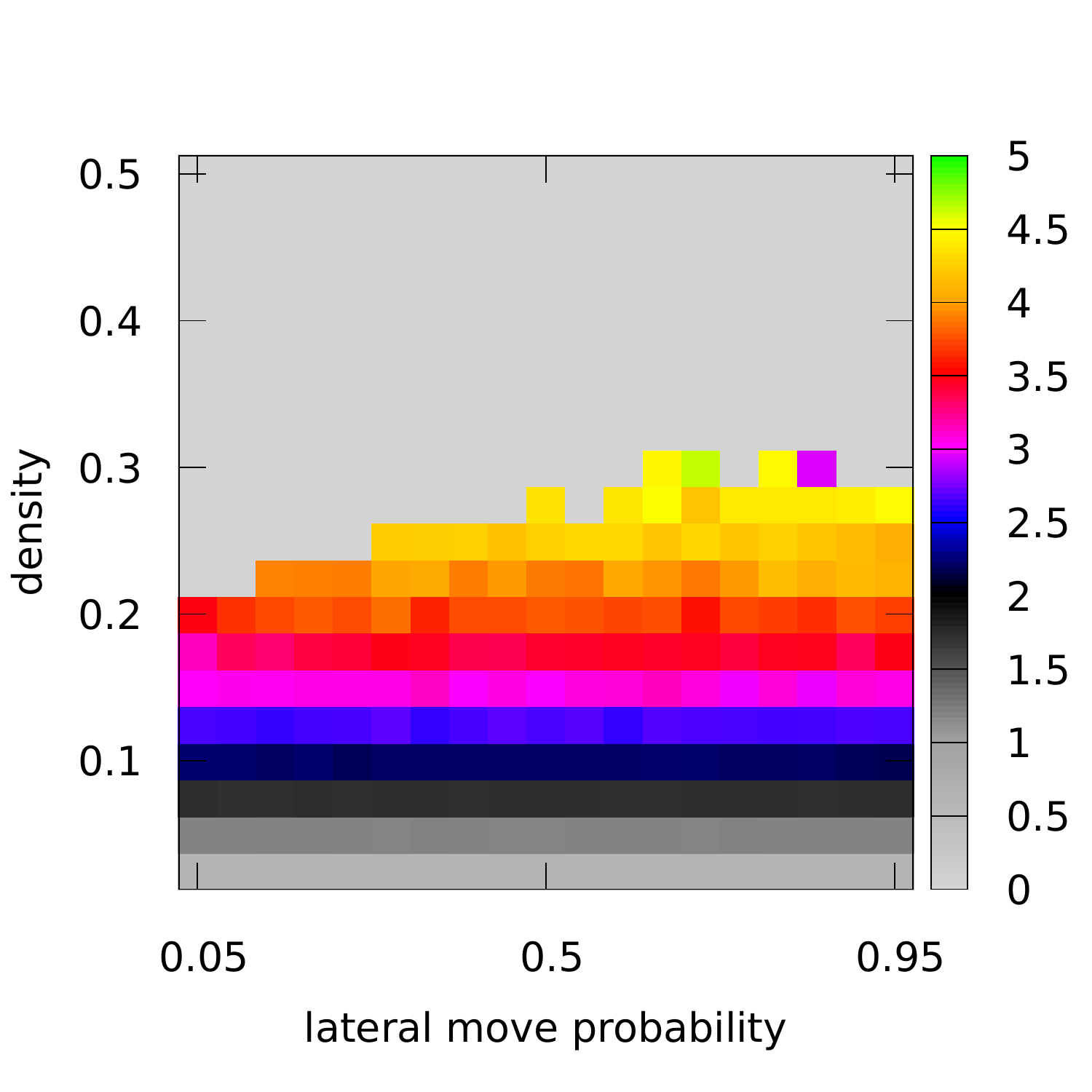}
}
\put(120,0)
{
 \includegraphics[width=.24\textwidth,height=.24\textwidth]{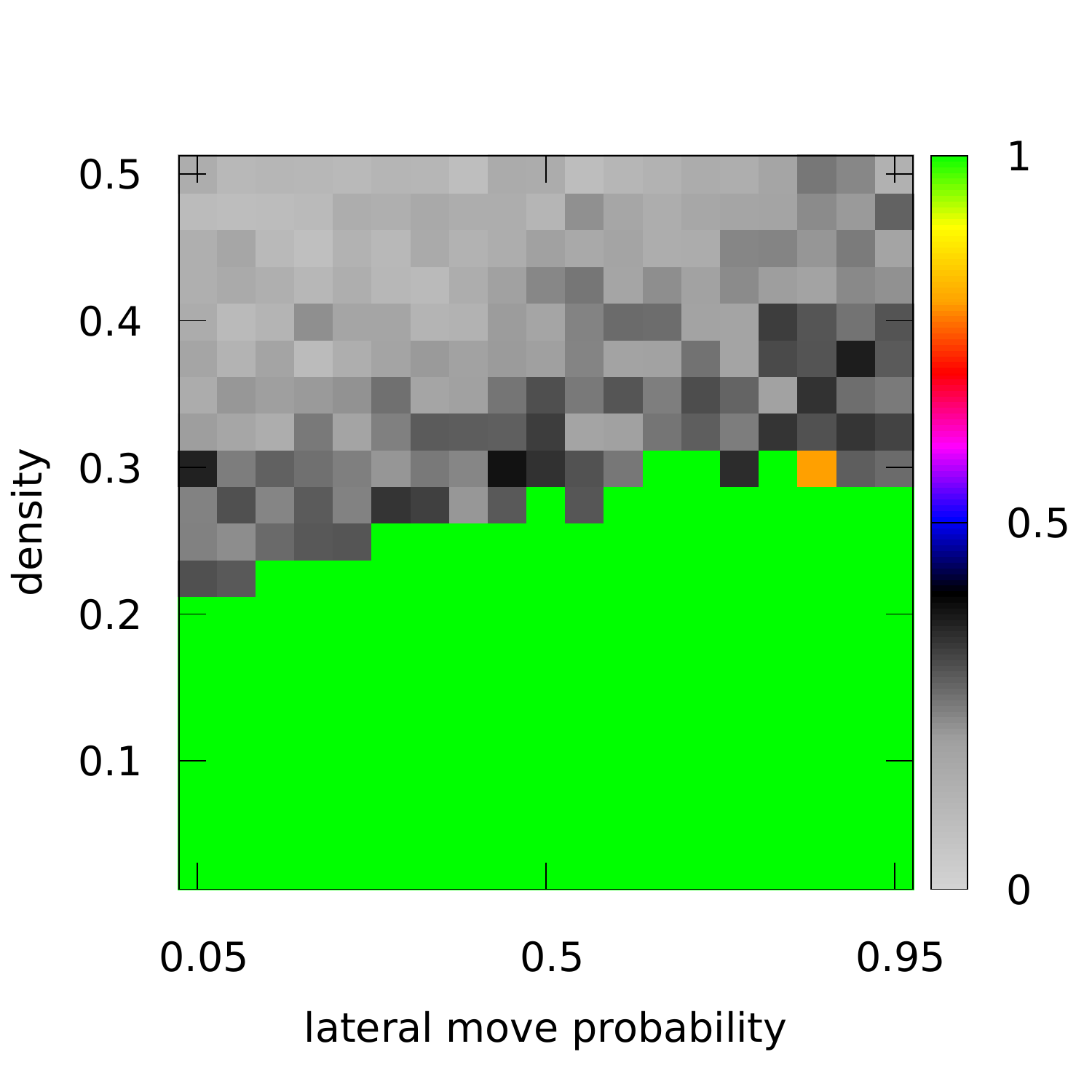}
}
\put(240,0)
{
 \includegraphics[width=.24\textwidth,height=.24\textwidth]{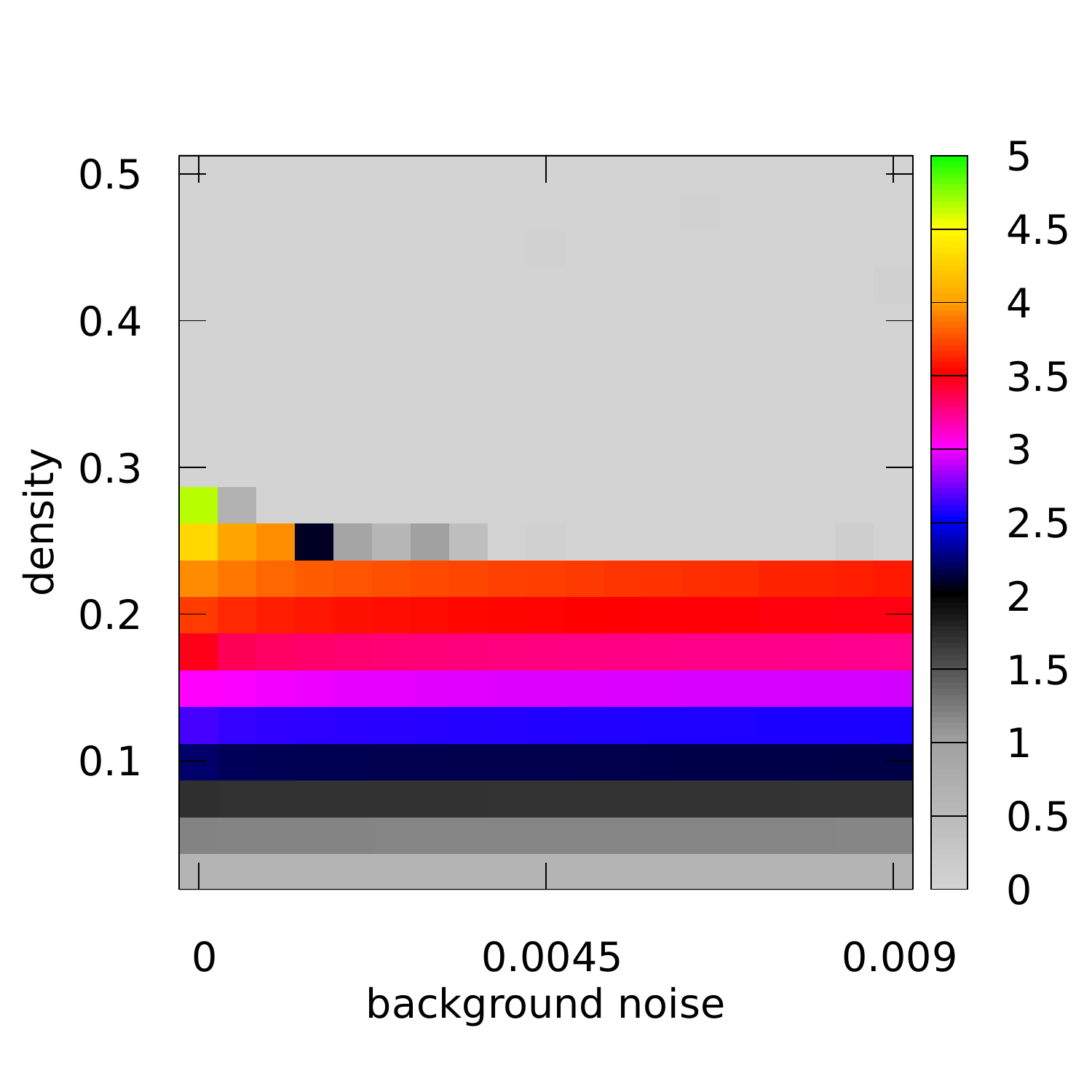}
}
\put(360,0)
{
 \includegraphics[width=.24\textwidth,height=.24\textwidth]{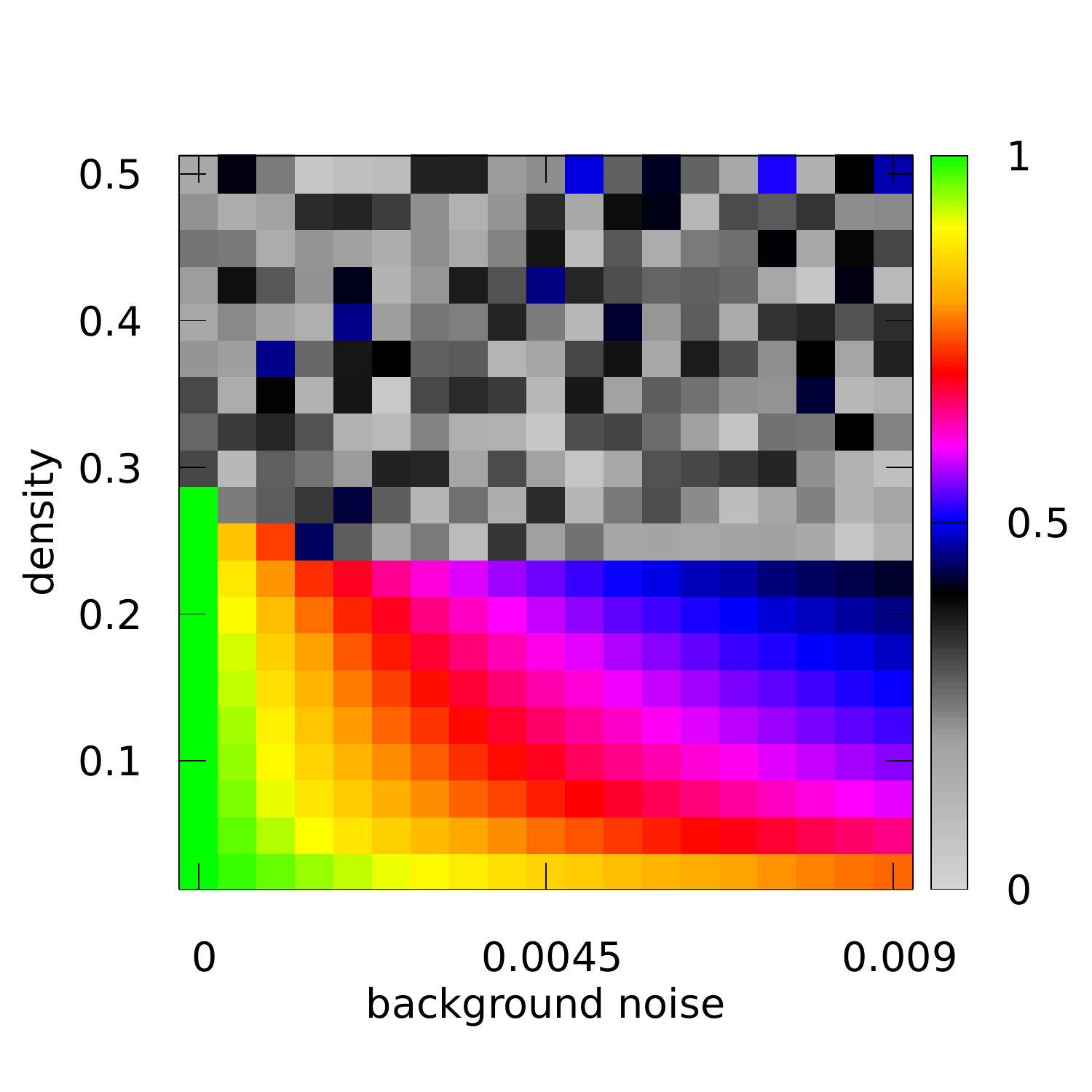}
}
\end{picture}
\caption{Scatter plot of the current (first and third panel) 
and order parameter $\phi$ (second and fourth panel)
for the strict periodic dynamics. 
In the first two panels $H=5$ and $r=0$. 
In the third and fourth panels $H=5$ and $h=0.71$.
}
\label{fig:n009} 
\end{figure}

We have repeated the simulations shown in the first two panels 
of Figure~\ref{fig:n005} and in the third panel of Figures~\ref{fig:n007} 
and \ref{fig:n008}. Our results, now plotted in Figure~\ref{fig:n009}, 
do not show new features with respect to what has been discussed above. 

\end{document}